\documentclass[trackchanges,twocolumn]{aastex7}

\newcommand{\Teff}{$T_\mathrm{eff}$}                 
\newcommand{\logg}{$\log(g)$} 
\newcommand{\err}{\,$\pm$\,}
\newcommand{\mum}{$\mu$m~}

\usepackage{CJK}

\begin{document}
\begin{CJK}{UTF8}{gbsn}
\title{Spectroscopic Monitoring of Metal Lines in Polluted White Dwarfs\footnote{based on observations made with the Southern African Large Telescope (SALT), 6.5m Magellan Telescope, and \textit{Hubble Space Telescope}.}}

\author[orcid=0000-0002-3553-9474,sname='L. K. Rogers']{Laura K. Rogers}
\affiliation{NOIRLab, 950 N Cherry Ave, Tucson, AZ, 85719, USA}
\email[show]{laura.rogers@noirlab.edu}  

\author[orcid=0000-0003-0155-2539]{Michael M. Shara} 
\affiliation{Department of Astrophysics, American Museum of Natural History, New York, NY, USA}
\email{mshara@amnh.org}

\author[orcid=0000-0002-8070-1901]{Amy Bonsor} 
\affiliation{Institute of Astronomy, University of Cambridge, Madingley Road, Cambridge CB3 0HA, UK}
\email{abonsor@ast.cam.ac.uk}

\author[orcid=0000-0002-8808-4282]{Siyi Xu (许\CJKfamily{bsmi}偲\CJKfamily{gbsn}艺)} 
\affiliation{NOIRLab, 950 N Cherry Ave, Tucson, AZ, 85719, USA}
\email{siyi.xu@noirlab.edu}

\author[orcid=0000-0002-3307-1062]{\'Erika Le Bourdais} 
\affiliation{Trottier Institute for Research on Exoplanets, D\'epartement de physique, Universit\'e de Montr\'eal, 1375 Ave. Th\'er\`ese-Lavoie-Roux Montr\'eal, QC H2V 0B3, Canada}
\email{erika.le.bourdais@umontreal.ca}
\affiliation{Centre de recherche en astrophysique du Qu\'ebec, Montr\'eal, QC, Canada}
\affiliation{Observatoire du Mont-M\'egantic, Montr\'eal, QC, Canada}
\author[orcid=0000-0003-4609-4500]{Patrick Dufour} 
\affiliation{Trottier Institute for Research on Exoplanets, D\'epartement de physique, Universit\'e de Montr\'eal, 1375 Ave. Th\'er\`ese-Lavoie-Roux Montr\'eal, QC H2V 0B3, Canada}
\affiliation{Centre de recherche en astrophysique du Qu\'ebec, Montr\'eal, QC, Canada}
\affiliation{Observatoire du Mont-M\'egantic, Montr\'eal, QC, Canada}
\email{patrick.dufour@umontreal.ca}

\author[orcid=0000-0002-1783-8817]{John Debes} 
\affiliation{Space Telescope Science Institute, Baltimore, MD, USA}
\email{debes@stsci.edu}

\author[orcid=0009-0008-0180-2554]{Omri Nolan} 
\affiliation{Institute of Astronomy, University of Cambridge, Madingley Road, Cambridge CB3 0HA, UK}
\email{omrinolan1@gmail.com}

\author[orcid=0000-0002-5775-2866]{Ted von Hippel} 
\affiliation{Department of Physical Sciences, Embry-Riddle Aeronautical University, Daytona Beach, FL 32114, USA}
\email{ted.vonhippel@erau.edu}

\author[orcid=0000-0003-2852-268X]{Erik Dennihy} 
\affiliation{NOIRLab, 950 N Cherry Ave, Tucson, AZ, 85719, USA}
\email{erik.dennihy@noirlab.edu}

\author[orcid=0000-0002-5470-3962]{Simon Hodgkin} 
\affiliation{Institute of Astronomy, University of Cambridge, Madingley Road, Cambridge CB3 0HA, UK}
\email{sth@ast.cam.ac.uk}

\author[orcid=0000-0001-6515-9854]{Andrew Swan} 
\affiliation{Department of Physics, University of Warwick, Coventry, CV4 7AL, UK}
\email{andrew.swan@warwick.ac.uk}

\author[orcid=0000-0003-4903-567X]{Mariona Badenas-Agusti} 
\affiliation{Institute of Astronomy, University of Cambridge, Madingley Road, Cambridge CB3 0HA, UK}
\email{mb2716@cam.ac.uk}

\author[orcid=0000-0001-9064-5598]{Mark C. Wyatt} 
\affiliation{Institute of Astronomy, University of Cambridge, Madingley Road, Cambridge CB3 0HA, UK}
\email{wyatt@ast.cam.ac.uk}

\author[orcid=0000-0001-7296-3533]{Tim Cunningham} 
\affiliation{Center for Astrophysics | Harvard \& Smithsonian, 60 Garden St, Cambridge, MA 02138, USA}
\email{timothy.cunningham@cfa.harvard.edu}

\begin{abstract}

The disruption and accretion of planetary material onto white dwarfs is expected to be inherently dynamic and stochastic, potentially driving variability in the accretion rate and therefore the shape and depth of the photospheric metal absorption lines. This paper presents an 18-year optical spectroscopic monitoring campaign of five warm (11,000--23,000\,K) polluted white dwarfs with sinking timescales of days--months, observed using Magellan/MIKE and SALT/HRS to directly test this prediction. At four of the five systems, no statistically significant variability is detected over baselines of 15--18 years corresponding to hundreds to thousands of diffusion timescales, with inferred accretion rates stable to within 15--30\% (1\,$\sigma$) showing remarkably stable accretion on decadal timescales. This implies that either the processes maintaining the accretion of the disrupted planetary material are stable on the same timescales, or that currently uncharacterized photospheric processes act to smooth observable abundance variations on these timescales.  The one exception, WD\,0106$-328$, shows statistically significant variability in the 4481\,\AA\ \ion{Mg}{2} doublet from the ground-based data. Yet no significant equivalent width or abundance changes are seen between two \textit{Hubble Space Telescope} ultraviolet spectra taken in 2016 and 2025, despite probing a larger set of transitions. This may imply that the ground-based observations witnessed a stochastic excursion from a stable baseline accretion rate, rather than a sustained change in the bulk accretion rate.

\end{abstract}



\section{Introduction}\label{sec:intro}

In recent years, evidence for the presence of planetary systems around white dwarf stars has grown substantially. While only a handful of planets have been directly detected orbiting white dwarfs \citep[e.g.][]{vanderburg2020giant,blackman2021jovian}, approximately 25--50\% of white dwarfs show indirect signatures of planetary material polluting their atmospheres \citep[e.g.][]{zuckerman2003metal, zuckerman2010ancient, koester2014frequency,OuldRouis2024constraints}. 

Polluted white dwarfs are thought to arise from the accretion of planetary debris that originates from the tidal disruption of planetesimals. These planetesimals are dynamically scattered onto star-grazing orbits, where they tidally disrupt, the resulting debris migrates inwards, sublimates, and ultimately accretes onto the white dwarf \citep{debes2002there, jura2003tidally,veras2014formation,brouwers2022road}. The accreted material typically has masses comparable to asteroids in the Solar System \citep{harrison2018polluted,turner2020modelling}. White dwarfs have strong surface gravities that result in heavy elements sinking out of their atmosphere on timescales far shorter than their cooling ages, so the continued presence of heavy metals in the photospheres requires ongoing or recent accretion from circumstellar material \citep{koester2009accretion}. Observations of close-in, disintegrating transiting planetesimals \citep[e.g.][]{vanderburg2015disintegrating,vanderbosch2021recurring}, and close-in circumstellar dust and gas disks further supports this model. Infrared excesses reveal dusty disks in 1.5--4\% of white dwarfs, with more than 100 such systems known \citep[e.g.][]{becklin2005dusty,wilson2019unbiased,lai2021infrared}. In at least 22 of these systems, double-peak emission lines reveal the presence of gas co-located with the dust \citep{gaensicke2006gaseous, gansicke2007sdss, gansicke2008sdss, melis2010echoes, farihi2012trio, melis2012gaseous, brinkworth2012spitzer, debes2012detection, dennihy2020five,melis2020serendipitous,gentile2020white,Bhattacharjee2025ZTF}. Furthermore, circumstellar gas in absorption has been seen from optical spectroscopy for a handful of white dwarfs \citep[e.g.][]{debes2012detection,steele2021characterization}, and from ultraviolet spectroscopy from \ion{Si}{4} lines \citep[e.g.][]{gansicke2012chemical,Zuckerman2026High}. 

These circumstellar environments are highly dynamic and variable. The first evidence of variable dust emission was for WD\,J0959$-$0200, where the fluxes in the 3.6\,{\textmu}m and 4.5\,{\textmu}m \textit{Spitzer} Infrared Array Camera channels dropped by approximately 35\,\% within a year \citep{xu2014drop}. Surveys have since revealed that the majority of polluted white dwarfs hosting dust disks exhibit mid-infrared variability over year long timescales \citep{swan2019most,swan2020dust,swan2021collisions,Noor2025Activity}. In contrast, near-infrared ($<2.5$\,{\textmu}m) variability is rarer with only a few cases reported \citep{xu2014drop,xu2018infrared,rogers2020near}. The largest variations in dust emission are typically found in systems that host both dusty and gaseous disks \citep{swan2020dust,Guidry2024using}, and recent work has demonstrated a direct correlation between dust and gas emission variability \citep{Rogers2025simultaneous}. More than half of the known white dwarfs with gaseous emission disks exhibit variation in the morphology and/or equivalent width of their lines over timescales from weeks to years. These variations are attributed to dynamical processes such as disk precession or the orbiting of planetesimals within the disk \citep{wilson2014variable,wilson2015composition,manser2015doppler, manser2016another, dennihy2018rapid,manser2019planetesimal,dennihy2020five,gentile2020white,melis2020serendipitous,Rogers2025simultaneous}. In addition, observations of transiting debris around white dwarfs show variability in the shapes, depths and timescales of the transits which can evolve on periods as short as days \citep[e.g.][]{vanderburg2015disintegrating,Gansicke2016highspeed,vanderbosch2021recurring,Farihi2022relentless} and their activity level can change over years \citep{Aungwerojwit2024longterm}.

Despite extensive evidence for variability in circumstellar gas and dust, there are few robust detections of variability in the photospheric metal lines of white dwarfs. Early spectroscopic monitoring of the polluted white dwarf G29-38 reported an increase of $\sim$70 percent in the equivalent width of the Ca\,\textsc{ii} K line over 3 years \citep{von2007discovery}. However, follow-up work by \citet{debes2008second} found no evidence for changes in the equivalent width of the Ca\,\textsc{ii} K line or in several other spectral lines, and suggested that earlier variability may be an artifact of differing spectral resolutions. \citet{rogers_2022} reported tentative variability over a 10 year period in the \ion{Mg}{2} 4481\,\AA~line for WD\,0106$-$328, and additional optical data reported in \citet{Farihi2026accretion} found changes in the Ca\,\textsc{ii} K and \ion{Mg}{2} lines. Papers studying time-series observations of circumstellar dust or gas around individual white dwarfs have additional reported no variability in photospheric lines  \citep{wilson2014variable,farihi2018dust,Johnson2022unusual}. To date, no systematic investigations have been carried out across a sample of white dwarfs to search for variability in metal line strengths and place constraints on the amplitude or timescales of accretion rate changes.

This paper reports a spectroscopic monitoring campaign of five warm hydrogen dominated polluted white dwarfs (spectral type DAZ) using two medium resolution spectrographs, and follow-up \textit{Hubble Space Telescope} (\textit{HST}) ultraviolet spectra of WD\,0106$-$328, due to optical variability, as outlined above. Warm DAZ white dwarfs have accretion rates on the order of days, therefore enabling the search for variability in the accretion rate using spectroscopic monitoring. Each white dwarf was observed over a total baseline of 15--18 years with between 18--47 datasets obtained, covering timescales of days, months and years and sampling hundreds of sinking timescales, depending on the white dwarf properties. Section \ref{Observations} reports the target selection, optical and ultraviolet spectroscopic observations of the selected targets, and data reductions and processing. Section \ref{Analysis} discusses the methods used to determine the equivalent width over time, with results reported in Section \ref{Results}. Finally Sections \ref{Discussion} and \ref{Conclusions} discuss and conclude the results from the spectroscopic monitoring survey highlighting implications for accretion onto white dwarf surfaces.

\section{Observations and Data Reduction} \label{Observations}

\subsection{Target Selection}

\begin{deluxetable*}{llllll}
\tablecaption{The sample of polluted white dwarfs used in this study highlighting their positions and distances from \textit{Gaia} \citep{GaiaCollaboration2016GAIA,GaiaCollaboration2023GaiaDR3}, white dwarf parameters (\Teff\ and \logg) derived from spectroscopic fits from previous works, mass (M$_{\mathrm{WD}}$), radius and log(q) (q = log$_{10}(M_{\mathrm{CVZ}}/M_{\mathrm{WD}})$ the ratio of the mass of the convection zone to the total white dwarf mass) derived using evolutionary models from \citet{Bedard2020spectral}, pollution abundances, and sinking timescales from \citet{koester2020new}. The dust row highlights whether they are confirmed to have an infrared excess from circumstellar dust, and the variable dust marks whether they were found to have variability in both 3.6 and 4.5\,\mum Spitzer bands above 3\,$\sigma$ in \citet{swan2020dust} and \citet{Noor2025Activity}.}
\tablehead{
\colhead{WD Name} & \colhead{WD\,0106$-$328} & \colhead{WD\,0408$-$041} & \colhead{WD\,1457$-$086} & \colhead{WD\,1929+011} & \colhead{WD\,2326+049} }
\startdata
Other name & HE\,0106$-$3253 & GD\,56 & ... & GALEX\,J193156.8+011745 & G29-38 \\ 
RA (J2000) & 01:08:36.035 & 04:11:02.169 & 14:59:52.990 & 19:31:56.932 & 23:28:47.637 \\
DEC (J2000) & $-$32:37:43.377 & 03:58:22.587 & $-$08:49:29.646 & +01:17:44.107 & 05:14:54.235 \\
Distance (pc) & 68.9 & 71.6 & 104.2 & 53.3 & 17.5 \\
\Teff (K) & 17350\,$^\alpha$ & 15300\,$^{\beta}$ & 22200\,$^\beta$ & 21200\,$^\gamma$ & 11800\,$^{\delta}$ \\
\logg & 8.12\,$^\alpha$ & 8.09\,$^{\beta}$ & 7.99\,$^\beta$ & 7.91\,$^\gamma$ & 8.40\,$^{\delta}$\\
Mass (M$_{\odot}$) & 0.69 & 0.67 & 0.62 & 0.58 & 0.86 \\
Radius (R$_{\odot}$) & 0.012 & 0.0122 & 0.0132 & 0.0140 & 0.0097 \\
log(q) & $-$16.71 & $-$16.19 & $-$15.78 & $-$16.00 & $-$14.00 \\
Dust? & Yes$^1$ & Yes$^2$ & -\,$^*$ &  Yes$^3$ &  Yes$^4$ \\
Variable Dust? & Yes & Yes & - &  No &  Yes \\
log(Ca/H) & $-$5.93$\pm$0.11\,$^a$ & $-$6.86$\pm$0.20\,$^a$ & $-$6.23$\pm$0.20\,$^a$ & $-$5.83$\pm$0.10\,$^b$ & $-$6.58$\pm$0.12\,$^c$  \\
log(Mg/H) & $-$5.57$\pm$0.20\,$^a$ & $-$5.55$\pm$0.20\,$^a$ & $-$5.47$\pm$0.20\,$^a$ & $-$4.10$\pm$0.10\,$^b$ & $-$5.77$\pm$0.13\,$^c$ \\
$\tau_{\textrm{Ca}}$ (d) & 1.13 & 2.75 & 5.50 & 4.39 & 80.45 \\
$\tau_{\textrm{Mg}}$ (d) & 1.83 & 4.58 & 9.59 & 7.48 & 86.01 \\
\enddata
\label{tab:WD-sample}
\tablecomments{\textbf{White dwarf parameter references:} ($\alpha$) \citet{xu2019compositions}, ($\beta$) \citet{gianninas2011spectroscopic}, ($\gamma$)  \citet{gansicke2012chemical}, ($\delta$) \citet{xu2014elemental}.
      \\ \textbf{Dust references:} (1) \citet{farihi2010strengthening}, (2) \citet{jura2007externally}, (3) \citet{rocchetto2015frequency}, (4) \citet{zuckerman1987excess}. \\ \textbf{Abundance References:} (a) \citet{xu2019compositions}, (b) \citet{melis2011accretion}, (c) \citet{xu2014elemental}.
      \\$^*$ Although it was previously identified to have an infrared excess, \citet{dennihy2017wired} report that the photometry is contaminated.
      }
\end{deluxetable*}

Five white dwarfs were selected for this study based on their physical criteria (Table\,\ref{tab:WD-sample}). Firstly, all targets are warm (\Teff$>$10,000\,K) white dwarfs with the DAZ spectral type. DAZ white dwarfs have relatively short metal sinking timescales on the order of days--years \citep{koester2009accretion}, making them well suited to detect changes in the amount of metals in the white dwarfs' photospheres over the baseline of the observations (see Table\,\ref{tab:WD-sample}). In contrast, DBZ white dwarfs (helium dominated with metals) exhibit much longer sinking timescales on the order of 10$^2$ to 10$^6$ years, making them less sensitive to short-term variations. The selected targets also lie within a temperature range that avoids both extremes: they are not too cool (\Teff$<$10,000\,K) such that the sinking timescales of the metals become longer than the baseline of the observations, and are not too hot such that radiative levitation could become a dominant mechanism to bring metals into the photosphere ($>$\,20\,--\,25,000\,K) \citep{chayer1995radiative, chayer1995improved,barstow2014evidence}. For one of the hottest targets, WD\,1929+011, radiative levitation has been shown to make a negligible contribution compared to external accretion \citep{koester2014frequency}. Finally, the targets must exhibit strong atmospheric pollution with detections of at least Ca and Mg in their optical spectra, with abundances of [Ca/H]$\,>\,-$7.0 and [Mg/H]$\,>\,-$6.0, as shown in Table\,\ref{tab:WD-sample}. The brightest targets observable with HRS/SALT were then selected for the study. Four of the five white dwarfs show an infrared excess indicative of circumstellar dust, however, this was not a selection requirement. G29-38 was included in this program, as it has been the target of previous studies \citep{von2007discovery,debes2008second}. As outlined in the introduction, it was proposed that variations were the result of comparing equivalent widths from spectrographs with differing resolutions, therefore, this target is re-visited using long-term monitoring campaigns with two spectrographs.

\subsection{Spectroscopic observations}  

\subsubsection{HRS/SALT} \label{HRS}
Spectroscopic observations of the white dwarfs were taken with the High Resolution Spectrograph \citep[HRS, ][]{barnes2008optical,Bramall2010HRS,Bramall2012HRS,Crause2014HRS} mounted on The Southern African Large Telescope (SALT) at the Sutherland Observatory in South Africa \citep{Buckley2006SALT}. Observations spanned across 12 semesters from Semester 2017/1 to 2025/1 covering 8 years; see Table \ref{tab:WD-Obs} for a summary of the range of dates of observation, exposure times, and signal-to-noise ratios (SNR). The HRS is a dual-beam (3700--5500\,\AA\ \& 5500--8900\,\AA) fiber fed, echelle spectrograph. The LR mode was used to yield a resolution of $\sim$14,000. This was a queue based program with both high priority time, where up to three sets of spectra were taken consecutively for each white dwarf to obtain high SNR data, and low priority time, where one spectrum was obtained as a filler between other programs or during poor weather. Therefore, for each observing night, the exposure times and SNR varied. The data were reduced by the MIDAS pipeline which produced flat fielded, wavelength calibrated and merged spectra \citep{kniazev2016mn48,Kniazev2017SALT}. For the observations where multiple consecutive spectra were obtained, they were continuum normalized using a third order polynomial fit to a region of 10--15\,\AA\ around the spectral line and stacked weighted by their inverse variance. Experiments with higher order polynomials tended to over-fit the continuum and include the wings of the lines, systematically reducing the equivalent widths of the lines.

\subsubsection{MIKE/Magellan} \label{MIKE}

Spectra were taken between 2007--2011 on the MIKE spectrograph at the Magellan Clay Telescope at the Las Campanas Observatory in Chile \citep{bernstein2003mike} (see Table \ref{tab:WD-Obs} for details on the observations). The MIKE spectrograph is a double echelle spectrograph with two arms covering wavelengths of 3350--5000\,\AA\ (blue) and 4900--9500\,\AA\ (red). The data were obtained under varying atmospheric conditions but almost always using the $0.7\arcsec\times5\arcsec$ slit. This gave a spectral resolution of approximately 35,000 for the red arm and 46,000 for the blue arm. The spectra were binned at $2\times2$ (in the spatial and spectral direction) and used a slow readout speed. The data were reduced using the CarPy suite of MIKE reduction tools to produce calibrated spectra \citep{Kelson2000Carpy,Kelson2003Carpy}. The MIKE/Magellan spectra exhibit substantial overlap between consecutive echelle orders, such that many spectral features are present in multiple orders. In addition, the instrument sensitivity functions are more complex. Therefore, to continuum normalize the data, the local continuum around each line was modeled using a polynomial with order 3--5, with the polynomial order selected according to the Bayesian Information Criterion. The spectra were then continuum normalized and combined using inverse-variance weighting.

\begin{deluxetable}{cccccc}
\label{tab:WD-Obs}
\tablecaption{Summary of optical observation details. SNR determined from the continuum around the \ion{Mg}{2} line at 4481.13\,\AA.}
\tablehead{\colhead{WD Name} & \colhead{Telescope} & \colhead{Date range} & \colhead{N epochs}  & \colhead{Exp Time range (s)} & \colhead{SNR range} }
\startdata
WD\,0106$-$328 & MIKE & 2007-10-25\,--\,2022-08-31 & 7 & 1800\,--\,3600 & 20.0\,--\,46.1 \\
WD\,0106$-$328 & SALT & 2019-06-08\,--\,2025-09-30 & 20 & 1450\,--\,2900 & 5.0\,--\,33.0 \\
WD\,0408$-$041 & MIKE & 2007-10-25\,--\,2009-10-14 & 6 & 3600\,--\,4800 & 32.4\,--\,53.8\\
WD\,0408$-$041 & SALT & 2018-11-28\,--\,2025-08-13 & 18 & 2375\,--\,3550 & 6.0\,--\,21.6 \\
WD\,1457$-$086 & MIKE & 2008-03-21\,--\,2010-08-03 & 6 & 3580\,--\,3600 & 9.5\,--\,59.8\\
WD\,1457$-$086 & SALT & 2018-06-05\,--\,2025-08-19 & 12 & 2375\,--\,3790 & 2.0\,--\,16.8 \\
WD\,1929+011 & MIKE & 2010-06-16\,--\,2011-06-09 & 6 & 900\,--\,2687 & 20.2\,--\,69.3 \\
WD\,1929+011 & SALT & 2017-07-06\,--\,2025-08-01 & 39 & 1625\,--\,3550 & 8.5\,--\,51.0 \\
WD\,2326+049 & MIKE & 2007-10-25\,--\,2011-06-09 & 15 & 400\,--\,3600 & 51.5\,--\,186.2 \\
WD\,2326+049 & SALT & 2017-07-05\,--\,2025-07-27 & 32 & 1775\,--\,3050 & 26.9\,--\,91.4 \\
\enddata
\end{deluxetable}

\subsubsection{\textit{HST}/COS}

\textit{HST} far ultraviolet (FUV) spectroscopic observations of WD\,0106$-$328 were obtained on 2025-09-29 with an exposure time of 3623.328 seconds as part of the program 17819. These are compared to historic data taken on 2016-08-28 with an exposure time of 2215 seconds as part of the program 14597 and reported in \citet{Farihi2026accretion}. The two sets of data were taken using the G130M grating with a central wavelength of 1291\,\AA, this gives a total wavelength coverage of 1150--1430\,\AA~(with a 20\,\AA~gap between the two segments) covering numerous metal lines across this region of the UV. The outputted data products from the \textsc{calcos} reduction pipeline were used for subsequent analysis. The SNR per resolution element is 18.9 for the data taken in 2016 and 21.0 for the data taken in 2025 as calculated from the continuum around the C\,\textsc{ii} lines at 1334.530 and 1335.708\,\AA. All the {\it HST} data used in this paper can be found in MAST: \dataset[10.17909/cxpw-zn51]{http://dx.doi.org/10.17909/cxpw-zn51}.

To assess whether instrumental variability between the two COS datasets could affect the shape of depth of the metal lines, the COS G130M line spread functions were compared for the relevant lifetime positions. The 2016 dataset used lifetime positions 1 through 4, whereas, the 2025 data used lifetime positions 3 and 4. The mean line spread functions were calculated for each dataset using online tables\footnote{\url{https://www.stsci.edu/hst/instrumentation/cos/performance/spectral-resolution}} and compared at various wavelengths across the detectors (1154, 1259, 1300, 1400\,\AA). A Kolmogorov-Smirnov test showed that the profiles were statistically consistent, and so there are no line spread function differences that could induce large systematic errors on the metal line shape or depths.

\section{Variability Analysis} \label{Analysis}

Equivalent widths are most often used to quantify the strengths of spectral lines; this is a measure of the area of the spectral line relative to the continuum level. To derive equivalent widths, a Voigt profile was fitted to the spectral lines, Voigt profiles were selected as they account for Doppler and pressure broadening and therefore provide a more accurate representation of observed absorption line shapes for white dwarfs \citep{armstrong1967spectrum}. A Markov Chain Monte Carlo (MCMC) approach was implemented to explore the range of Voigt profiles which fit the fluxes and associated errors for each spectral line; this better quantifies the uncertainties. A six parameter Voigt profile was used, consisting of a straight line ($y=Mx + C$), to account for additional continuum variation, and a Voigt profile model. Voigt profiles are a convolution of the Gaussian and Lorentzian profiles. The Voigt profile does not have a defined equation but is related to the real part of the Faddeeva function, $w(z)$, with $z = \frac{x + i\gamma}{\sigma \sqrt{2}}$:
\begin{equation}
    V = \frac{\textrm{Re}[w(z)]}{\sigma \sqrt{2 \pi}},
\end{equation}
which gives the model to be fitted to the data as: 
\begin{equation}
    V = (Mx + C)  + A \frac{\textrm{Re}[w((x - x_0 + i\gamma)/\sigma \sqrt{2})]}{\sigma \sqrt{2 \pi}},
\end{equation}
where $\sigma$ is the standard deviation of the Gaussian profile, $\gamma$ is the half-width at half-maximum of the Lorentzian profile, $x_0$ is the wavelength value marking the center of the line, $A$ is the amplitude, and $M$ and $C$ define the slope and intercept of the straight line. The \textsc{emcee}\footnote{\url{https://emcee.readthedocs.io/en/stable}} python package was used to implement the MCMC sampling. 24 walkers were used to sample the posterior distributions of the parameters with 50,000 iterations each; 25,000 iterations were run and discarded for the burn-in phase. The chains were thinned by approximately the auto-correlation time, the number of iterations required for the walkers to `forget' their initial position, to ensure the parameters in the chain were independent. For each set of Voigt model parameters in the chain ($M$, $C$, $A$, $x_0$, $\sigma$, $\gamma$), the equivalent width was evaluated. The Voigt profile was integrated with bounds of $\pm$\,3 times the full-width at half-maximum of the Voigt model, and divided by the continuum to give the equivalent width for that set of parameters. From this chain, the median value of the equivalent width was calculated and the errors are quoted using the 16th and 84th percentiles. 

Accurate and precise continuum normalization is critical for the analysis. The continuum around the spectral line of interest was fitted for HRS/SALT and MIKE/Magellan as outlined in Sections \ref{HRS} and \ref{MIKE}. To quantify the systematic uncertainty associated with continuum normalization, the equivalent widths were recomputed using a high SNR spectrum while varying the continuum fitting procedure. The order of the polynomial was varied between 2--4 and the wavelength range between 5--15\,\AA\ either side of the line. These variations result in changes in the measured equivalent widths of 3.2\% for MIKE/Magellan and 1.5\% for HRS/SALT. These values represent lower limits on the true systematic uncertainty as they are derived from relatively high SNR data and do not fully capture additional uncertainties introduced in lower quality spectra or by other sources of systematic error, such as continuum placement biases and instrumental effects. To conservatively account for these effects, an uncertainty corresponding to twice the measured variation in equivalent width when propagating errors associated with continuum normalization is adopted. These errors were combined in quadrature with the 16th and 84th equivalent width percentiles from the MCMC Voigt fit. 

The equivalent width results were also checked against direct integration measurements on the continuum-normalized spectra. An initial Gaussian profile is fitted to each spectral feature to first determine the line centroid and width. This is used to determine the integration window, defined as $\pm$\,3 times the full-width at half-maximum around the line center. The line flux is computed using trapezoid integration within this window, without assuming any parametric line profile. Uncertainties are estimated by perturbing the flux at each wavelength using a Gaussian with width of the flux errors and recomputing the integrated flux. This bootstrapping method was repeated 100,000 times and the 16th and 84th percentiles of the resulting distribution were adopted as the 1\,$\sigma$ confidence intervals. 

G29-38 is a ZZ Ceti pulsator and therefore experiences periodic changes in brightness of between 1--3\%. These pulsations can result in equivalent widths that are discrepant from the mean equivalent width. Longer integration times \citep[$\gg$ 23 mins - approximately the maximum pulsation period,][]{Winget1990Whole,Kleinman1998Understanding,Uzundag2023Asteroseismological} average out the effects of the pulsations on the equivalent widths \citep{von2007discovery}. Therefore, for spectra taken on consecutive days, these are stacked and processed together to obtain a longer total exposure to gain more accurate measurements of equivalent width. Given the sinking timescales are much greater than this (Table \ref{tab:WD-sample}), this does not bias results. 

\section{Results} \label{Results}

\subsection{Equivalent Width}

\begin{figure*}
\centering
{
  \includegraphics[width=0.48\textwidth]{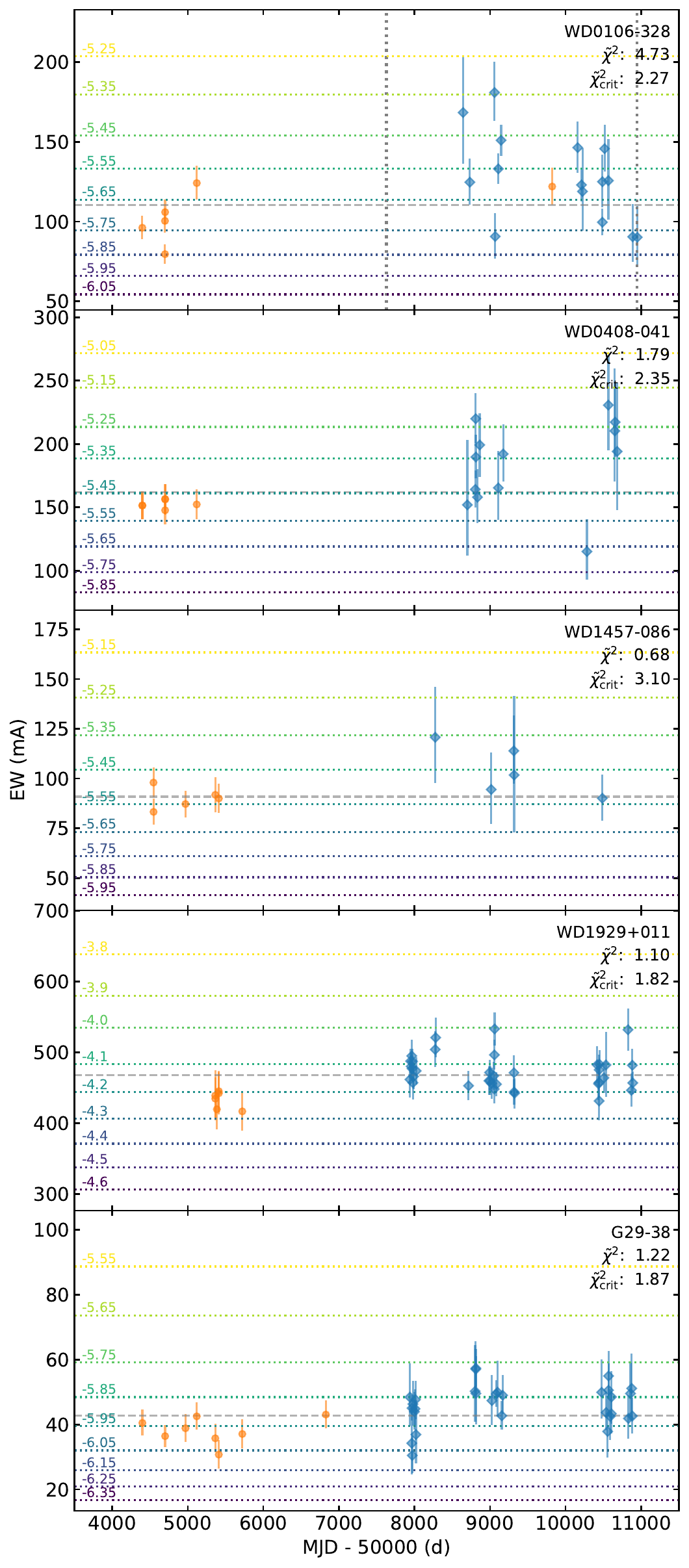}
}
\vspace{2mm}
{
  \includegraphics[width=0.47\textwidth]{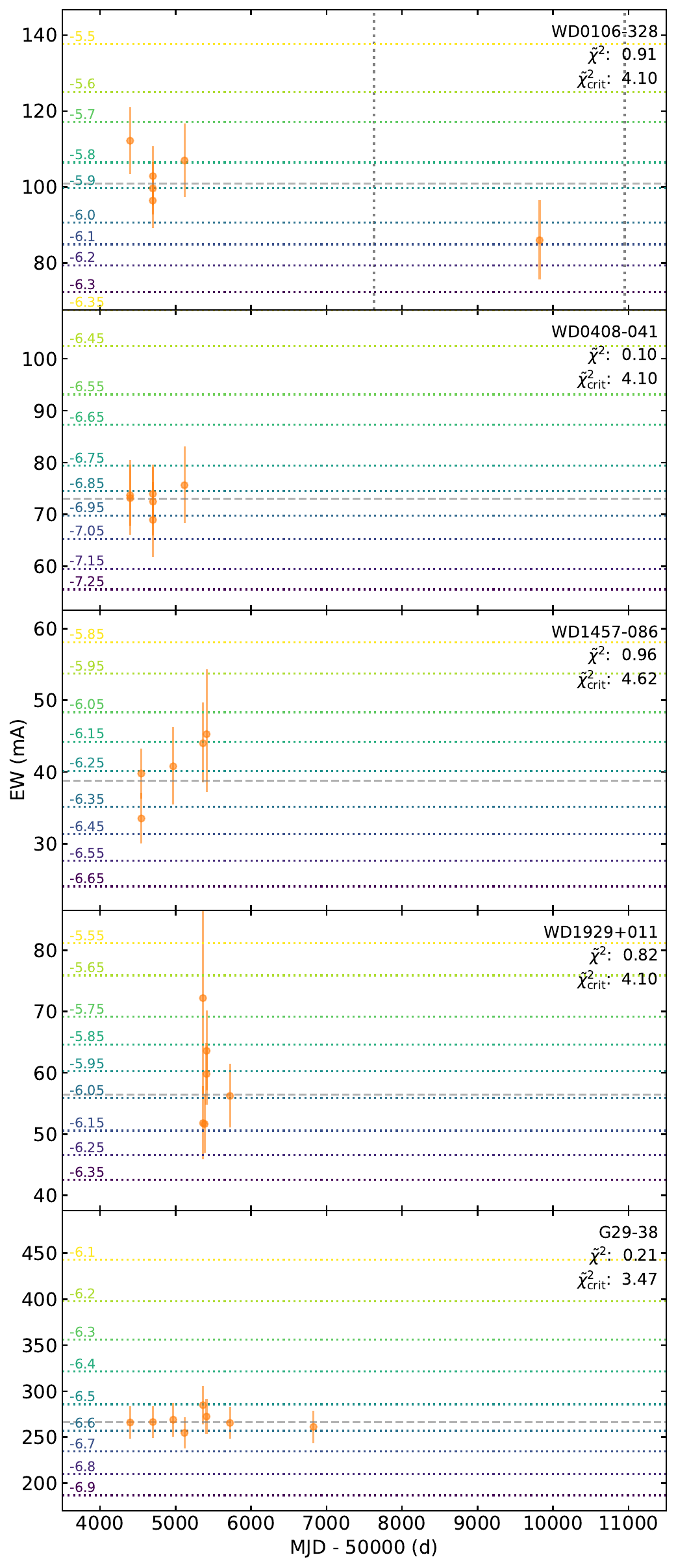}
}
\caption{The equivalent width of (left) the \ion{Mg}{2} doublet and (right) the Ca\,\textsc{ii} K line over time for each of the white dwarfs with MIKE/Magellan data marked as orange circles and HRS/SALT marked as blue diamonds for data with SNR$>$10 and lines detected at a 3\,$\sigma$ significance. The dashed line marks the weighted mean of the equivalent widths. The dotted horizontal lines mark the equivalent width for a spectral line with the quoted [Mg/H] and [Ca/H] abundances, with the y axis range spanning 1\,dex in [Mg/H] or [Ca/H] respectively. For WD\,0106$-$328 the vertical dotted lines mark the epochs in which the \textit{HST} data were taken.}
\label{fig:EW-Time}
\end{figure*}

\subsubsection{MIKE/Magellan and HRS/SALT} \label{results-m-s}
This work reports spectroscopic monitoring campaigns of five DAZ white dwarfs to investigate whether the mass of heavy elements in their photosphere changes over time. For cooler DAZ white dwarfs (below approximately 12,000\,K), such as G29-38, the strongest metal line is the Ca\,\textsc{ii} K line at 3933.66\,\AA, but for hotter DAZs the dominant optical metal line shifts to the \ion{Mg}{2} doublet at 4481.13\,\AA, as calcium becomes increasingly doubly ionized whilst magnesium remains predominantly in the \ion{Mg}{2} state. HRS/SALT has limited sensitivity around 3933.66\,\AA\ and therefore the Mg\,\textsc{ii} 4481.13\,\AA\ line is used as the primary tracer of variability in this study.  

Figure\,\ref{fig:EW-Time} (left) shows the equivalent width of the \ion{Mg}{2} 4481.13\,\AA\ line over time for the five white dwarfs in the sample. To test whether the equivalent widths are consistent with a constant value and therefore no variability, a horizontal line was fitted to the data and the goodness-of-fit was evaluated using the reduced chi squared statistic ($\tilde{\chi}^2 = \chi ^2 / N(\textrm{dof})$, where $N(\textrm{dof})$ is the number of degrees of freedom). A $\tilde{\chi}^2$ of 1 indicates that the constant model provides a good description of the data, $\tilde{\chi}^2$ $<$ 1 implies the model is over-fitting the data, $\tilde{\chi}^2$ $>$ 1 implies the model is under-fitting the data or the uncertainties are over-estimated, and $\tilde{\chi}^2$ $\gg$ 1 implies the model is a poor fit to the data. To determine whether the constant equivalent width model (the null hypothesis) can be rejected, the measured $\tilde{\chi}^2$ values are compared to critical values, $\chi ^2_{\textrm{crit}}$, corresponding to a confidence level of 99.9\%. If $\tilde{\chi} ^2 > \tilde{\chi} ^2_{\textrm{crit}}$ the null hypothesis is rejected and the equivalent width is considered variable.

Following this analysis, four of the white dwarfs in the sample show no statistically significant variability in the equivalent width of the \ion{Mg}{2} line over the monitoring baseline with MIKE/Magellan and SALT/HRS as shown in Fig.\,\ref{fig:EW-Time} (left) with $\tilde{\chi}^2$ and $\chi ^2_{\textrm{crit}}$ values reported in Table \ref{tab:WD-Acc-Rate}, and no statistically significant variability in the equivalent width of the \ion{Ca}{2} K line over the baseline of the MIKE/Magellan as shown in Fig.\,\ref{fig:EW-Time} (right). For these systems, the measured $\tilde{\chi}^2$ values are consistent with a constant equivalent width model at the 99.9\% confidence level. The dotted lines in Figure\,\ref{fig:EW-Time} show the abundance of Mg and Ca that would give the plotted equivalent width showing that these white dwarfs have equivalent widths that do not vary on average above $\pm0.09$\,dex (to 1\,$\sigma$) and $\pm0.27$\,dex (to 3\,$\sigma$). In contrast, from the top panel in Fig.\,\ref{fig:EW-Time} (left) there appears to be a significant offset between the equivalent widths measured from the MIKE/Magellan data and those measured from the first datasets of HRS/SALT for WD\,0106$-$328. Indeed the $\tilde{\chi} ^2$ value is high at 4.73, driven by the large equivalent widths measured for the HRS/SALT data around 9000 MJD--50000 days. The $\tilde{\chi} ^2_{\textrm{crit}}$ value is 2.27, and so the null hypothesis of a constant equivalent width can be rejected at the 99.9\% confidence level. 

\begin{figure}
    \centering
	\includegraphics[width=1.0\columnwidth]{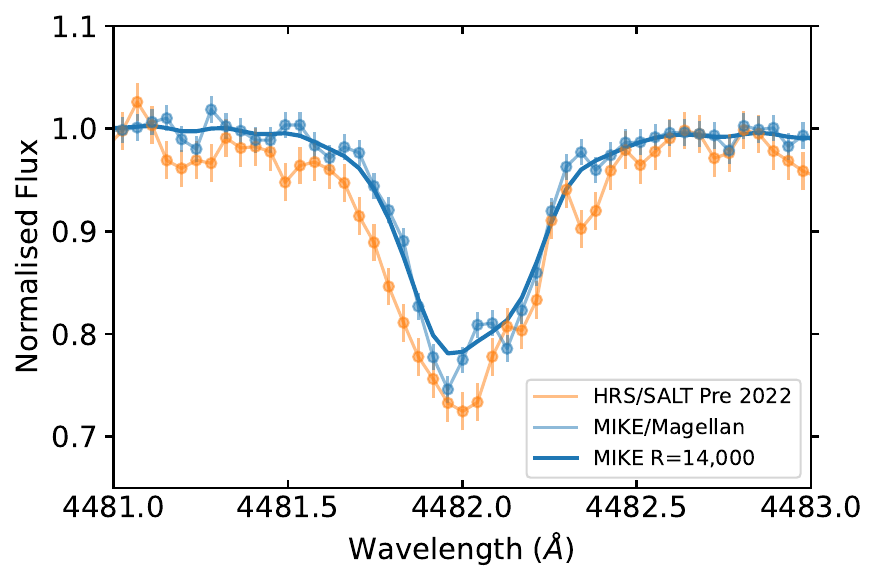}
    \caption{The stacked MIKE/Magellan and SALT/HRS data (pre-2022) for the \ion{Mg}{2} doublet for WD\,0106$-$328. The darker blue line is the MIKE/Magellan spectra interpolated and convolved to the resolution of HRS/SALT. }
    \label{fig:0106-stacked-res}
\end{figure}

To further investigate potential long term changes, the MIKE/Magellan spectra and the HRS/SALT spectra were stacked separately for each star to produce two higher SNR spectra separated by approximately a decade. The equivalent widths measured from these stacked spectra for \ion{Mg}{2} and other resolved spectral lines are reported in Table~\ref{tab:Multiple-Lines}. For four of the white dwarfs, all the lines are consistent within 1--2\,$\sigma$ indicating that the mass and abundances of accreted material in the photospheres of these four white dwarfs remain stable over the observed timescales. In contrast, WD\,0106$-$328  shows evidence for a 2.8\,$\sigma$ increase in equivalent width of the \ion{Mg}{2} doublet between the MIKE/Magellan the HRS/SALT stacked spectra, this increases to 4.4\,$\sigma$ if only the pre-2022 HRS/SALT data is used in the stack. This is also shown in Fig.\,\ref{fig:0106-stacked-res}. 

\begin{table}
	\centering
	\footnotesize
	\caption{The abundances derived from the COS/\textit{HST}  data for WD\,0106$-$328 for the 2016 and 2025 data. All abundances are consistent within 1.2\,$\sigma$.}
	\label{tab:HST-abundances}
	\begin{tabular}{ccc} 
		\hline
		[X/H] & 2016 & 2025 \\
		\hline	
		  O & $-$5.30$\pm$0.10 & $-$5.28$\pm$0.12 \\
        Si & $-$5.83$\pm$0.12 & $-$5.83$\pm$0.12 \\
        Fe & $-$5.43$\pm$0.10 & $-$5.26$\pm$0.10 \\
        Ni & $-$6.94$\pm$0.16 & $-$6.94$\pm$0.10 \\
		\hline
	\end{tabular}
\end{table}

\begin{table*}
	\centering
	\footnotesize
	\caption{The median equivalent width and average error for the stacked  MIKE/Magellan (M) and HRS/SALT (S) data for weaker lines for the white dwarfs.}
	\label{tab:Multiple-Lines}
	\begin{tabular}{cccccccccccccccc} 
		\hline
		Line & $\lambda _{\textrm{air}}$ & 
  \multicolumn{2}{ c }{WD\,0106 EW} & \multicolumn{2}{ c }{WD\,0408 EW} & \multicolumn{2}{ c }{WD\,1457 EW} &
  \multicolumn{2}{ c }{WD\,1929 EW} & \multicolumn{2}{ c }{G29-38 EW} \\
   & (\AA) & \multicolumn{2}{ c }{(m\AA)} & \multicolumn{2}{ c }{(m\AA)} & \multicolumn{2}{ c }{(m\AA)} & \multicolumn{2}{ c }{(m\AA)} & \multicolumn{2}{ c }{(m\AA)} \\
    & & M & S & M & S & M & S & M & S & M & S \\

		\hline	

        \ion{Ca}{1} & 4226.73 & - & - & - & - & - & - & - & - & 22.8$\pm$2.2 & 21.4$\pm$2.0 \\
        \ion{Mg}{2} & 4481.13 & 100.7$\pm$6.6 & 123.4$\pm$4.7 & 152.3$\pm$10.0 & 181.5$\pm$8.6 & 88.3$\pm$5.9 & 97.8$\pm$6.1  & 431.4$\pm$28.4 & 477.2$\pm$17.6 & 36.9$\pm$2.5 & 45.5$\pm$1.7  \\
        \ion{Fe}{2} & 4923.93 & - & - & - & - & - & -  & - & - & - & -  \\
        \ion{Fe}{2} & 5018.44 & 15.2$\pm$2.3 & 18.1$\pm$1.7 & - & - & - & - & 20.7$\pm$1.8 & 22.5$\pm$1.2 & - & - \\
        \ion{Fe}{2} & 5169.03 & 21.2$\pm$2.0 & 17.9$\pm$2.8 & - & - & - & - & 32.7$\pm$3.0 & 26.4$\pm$1.4 & - & -  \\
        \ion{Si}{2} & 5041.02 & - & - & - & - & - & - & 96.1$\pm$7.8 & 124.5$\pm$7.6 & - & -  \\
        \ion{Si}{2} & 5055.98 & - & - & - & - & - & - & 218.8$\pm$17.8 & 210.3$\pm$7.1 & - & -  \\
        \ion{Si}{2} & 5957.56 & - & - & - & - & - & - & 52.9$\pm$5.6 & 64.3$\pm$5.3 & - & -  \\
        \ion{Si}{2} & 5978.93 & - & - & - & - & - & - & 57.0$\pm$4.8 & 45.7$\pm$2.4 & - & -  \\
        \ion{Si}{2} & 6347.10 & - & - & - & - & - & - & 151.0$\pm$10.2 & 159.3$\pm$5.4 & - & -  \\
        \ion{Si}{2} & 6371.36 & - & - & - & - & - & - & 107.0$\pm$7.1 & 116.6$\pm$3.8 & - & -  \\
        \ion{Mg}{1} & 5172.68 & - & - & - & - & - & - & - & - & 19.3$\pm$1.5 & 23.6$\pm$1.2 \\
        \ion{Mg}{1} & 5183.60 & - & - & - & - & - & - & - & - & 36.1$\pm$2.3 & 35.2$\pm$2.2 \\
        \ion{Mg}{2} & 7877.05 & - & - & - & - & - & - & 60.6$\pm$19.1 & 75.5$\pm$5.8 & - & -  \\
        \ion{Mg}{2} & 7896.37 & - & - & - & - & - & - & 114.1$\pm$20.0 & 125.7$\pm$10.6 & - & -  \\
        \ion{O}{1} & 7771.94 & - & - & - & - & - & - & 44.2$\pm$4.6 & 44.0$\pm$2.3 & - & -  \\
        \ion{O}{1} & 7774.17 & - & - & - & - & - & - & 32.1$\pm$4.2 & 22.2$\pm$1.7 & - & -  \\
		\hline
	\end{tabular}
\end{table*}

\begin{figure}
    \centering
	\includegraphics[width=1.0\columnwidth]{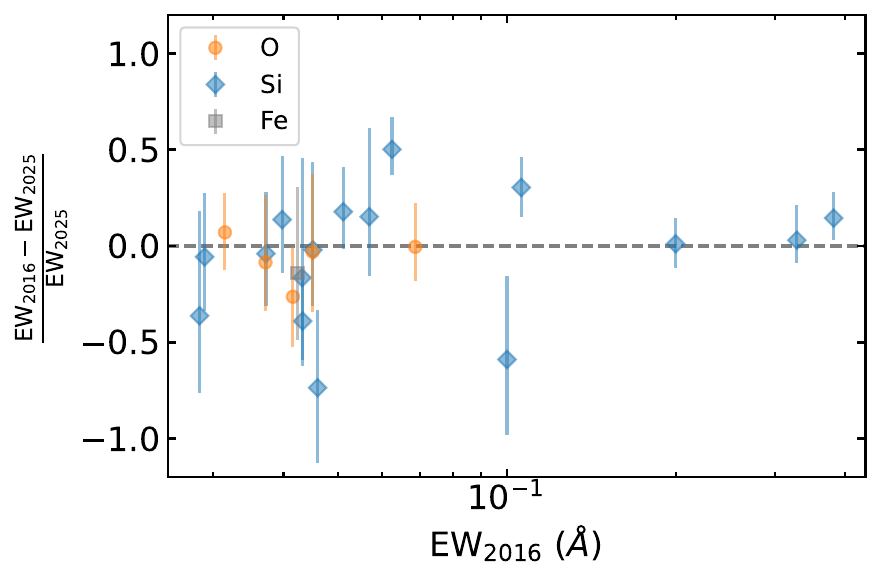}
    \caption{Comparison between the measured equivalent widths of the spectral lines between the 2016 and 2025 \textit{HST} data. Only lines detected at 3\,$\sigma$ in \textit{both} the 2016 and 2025 data are plotted.}
    \label{fig:0106-Hubble}
\end{figure}

\subsubsection{COS/\textit{HST} }

To further test whether the variability observed in the ground-based optical data reflects a genuine change in the accretion rate or bulk abundances for WD\,0106$-$328, ultraviolet spectra obtained with the \textit{HST} were analyzed using the same methods as reported in Section\,\ref{Analysis} and compared between the two epochs, 2016 and 2025. The equivalent widths of the Voigt profiles are given in Appendix A Table~\ref{tab:EW-HST-WD0106}. Only lines detected at the $>3\sigma$ level in both epochs were included in the comparison to ensure robust comparisons. 

For these lines, no statistically significant variation in equivalent width is detected at the $3\sigma$ level between the two \textit{HST} observations. This is illustrated in Figure \ref{fig:0106-Hubble} where the difference in equivalent width between the two epochs is shown. If the accretion rate changed, the measurements would preferentially lie on one side of zero, whereas, if the abundance of the material changed, different elements would be offset by different amounts. Instead, the measurements are consistent with being symmetrically distributed about zero within their uncertainties. No statistically significant differences are found for O, Fe, or Si between the 2016 and 2025 spectra, showing that the inferred atmospheric abundances were identical within error for these two epochs.

White dwarf models were fitted to the \textit{HST} data using white dwarf parameters from Table~\ref{tab:WD-sample} using the methods in \citet{dufour2012detailed}. The spectra were fitted in regions of 5--15\,\AA. Where an element was present in multiple regions, the average abundance was derived. The uncertainties have two contributions: the spread in abundances derived for a particular element when present in multiple regions and the error from the equivalent width, following \citet{Rogers2023sevenI}. The resulting abundances for the 2016 and 2025 datasets are reported in Table~\ref{tab:HST-abundances}, and the corresponding fits to  spectral regions are shown in Fig.\,\ref{fig:0106-Hubble}. Within the uncertainties, the measured abundances are consistent between the two epochs, with all elements agreeing within $1.2\,\sigma$. This result supports the equivalent width analysis, indicating that no statistically significant change in the atmospheric composition of the accreted material was detected between the two \textit{HST} observations. Table~\ref{tab:WD-Acc-Rate} reports the 1\,$\sigma$ abundance range from the ground-based observations, and this is greater than the 1\,$\sigma$ abundance range from the \textit{HST} observations reported in Table~\ref{tab:HST-abundances}. Therefore the \textit{HST} observations are more constraining, and if WD\,0106$-$328 were varying, the \textit{HST} observations were sensitive enough to detect these changes. These results suggest that if the accretion rate did vary, the change either occurred prior to the 2016 \textit{HST} observations or varied between 2016 and 2025 before returning to a similar level by the time of the later epoch. 

\begin{figure}
    \centering
	\includegraphics[width=1.0\columnwidth]{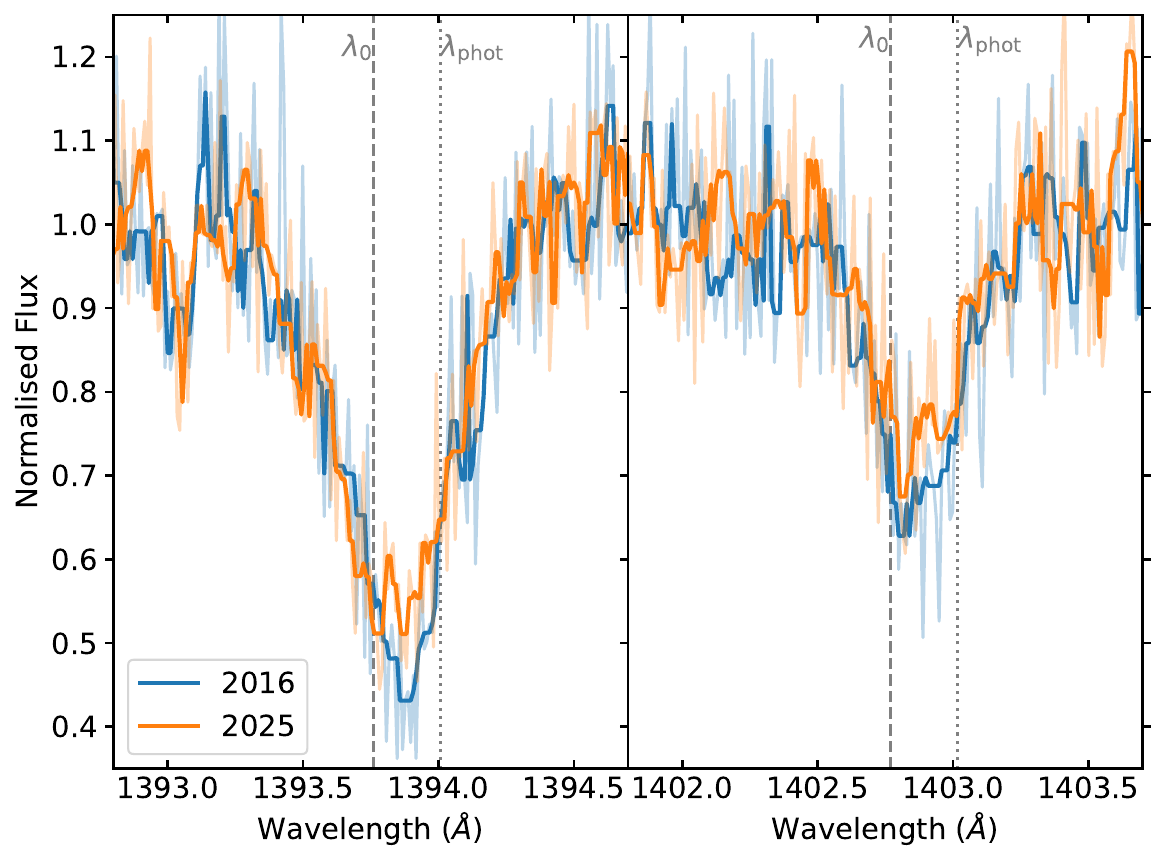}
    \caption{A comparison between the 2016 (blue) and 2025 (orange) \textit{HST} data of the two \ion{Si}{4} lines. The data are normalized and smoothed with a box size of 5 for clarity with the raw data faded in the background. $\lambda_{\textrm{0}}$ marks the vacuum wavelength of the \ion{Si}{4} lines and $\lambda_{\textrm{phot}}$ marks where the photospheric lines would lie, there is a clear offset showing these are likely circumstellar features. There is a tentative decrease in the line between 2016 and 2025.}
    \label{fig:Si-IV}
\end{figure}

WD\,0106$-$328 exhibits \ion{Si}{4} absorption features that are offset from the photospheric radial velocity by approximately $-$31\,km\,s$^{-1}$ and $-$36\,km\,s$^{-1}$ in the 2016 and 2025 data, respectively. These velocities are comparable to the expected gravitational redshift of the white dwarf \citep[36.5\,km\,s$^{-1}$, ][]{Bedard2020spectral}, indicating that the absorbing material is not photospheric in origin and is likely circumstellar. Such velocity offsets are characteristic of circumstellar gas observed in polluted white dwarfs and are consistent with similar detections reported in a growing sample of systems \citep{gansicke2012chemical,Rogers2023sevenI,Rogers2025Silicate,Zuckerman2026High}. As shown in Figure\,\ref{fig:Si-IV}, both \ion{Si}{4} lines appear stronger in the 2016 spectrum compared to the 2025 observations. While this difference does not reach a 3\,$\sigma$ significance (2\,$\sigma$ combined), follow up data may confirm that this white dwarf has variable circumstellar gas, analogous to other white dwarfs with dynamically evolving gaseous disks \citep{Zuckerman2026High}. 

\begin{figure}
    \centering
	\includegraphics[width=0.83\columnwidth]{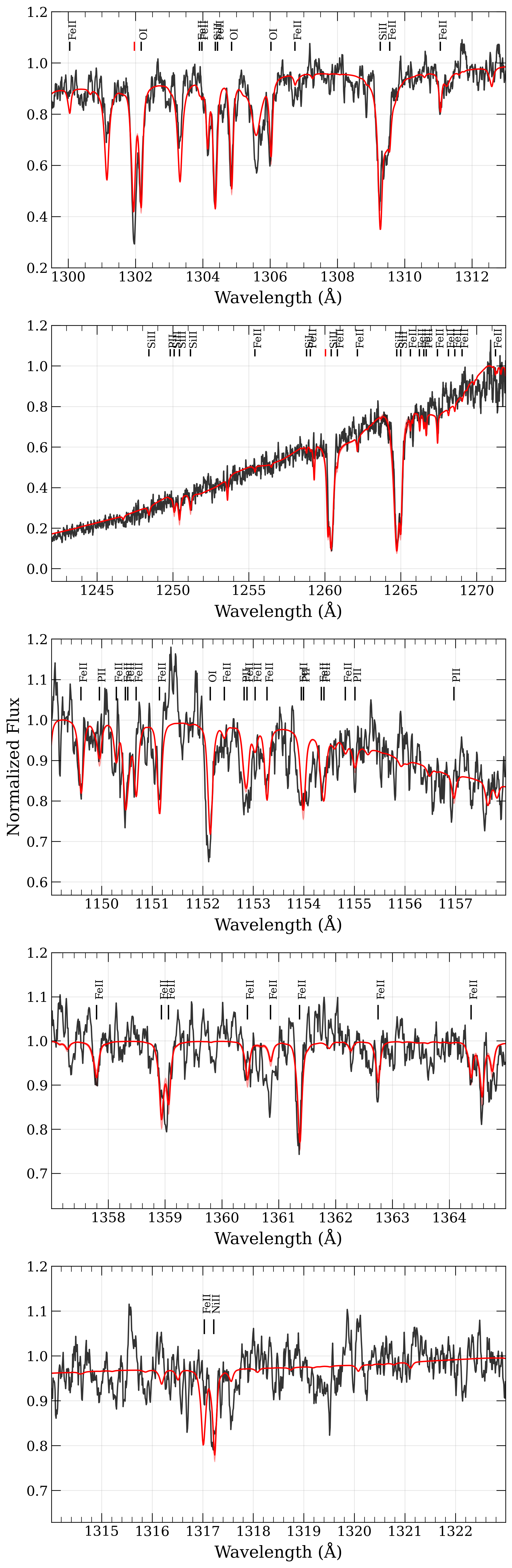}
    \caption{White dwarf model fits (red line) to the 2025 COS/\textit{HST} data for WD\,0106$-328$. Spectral lines are labeled, and those with an interstellar medium component have an additional red dash at the position of the interstellar medium lines.}
    \label{fig:Hubble-fits}
\end{figure}

\subsection{Radial Velocity constraints}

Fitting of individual spectral lines naturally leads to precise derivations of line centroids and hence radial velocities, enabling searches for stellar and, where precision permits, sub-stellar companions. Neither SALT/HRS nor MIKE/Magellan are optimized for ultra-precise radial velocity work, as both lack vacuum enclosures and strict temperature stabilization, limiting their long-term instrument stability. For SALT/HRS in Low Resolution Mode (LRS), the instrument stability has been quantified at 0.2\,km\,s$^{-1}$ for both red and blue channels, with a maximum of 0.5\,km\,s$^{-1}$ if there has not been a calibration within 50 days of the measurements \footnote{\url{https://www.saao.ac.za/~akniazev/pub/HRS_MIDAS/HRS_stability.pdf}}. Magellan/MIKE has a demonstrated performance at the level of 3\,m\,s$^{-1}$ precision under optimal conditions for bright, spectral line-rich main sequence stars \footnote{\url{https://ebps.carnegiescience.edu/telescopes/magellan-telescope}}); however, this level of precision is not generally achievable for white dwarf spectra due to their broader and fewer spectral features and typically lower SNRs \citep{Rogers2024WD0141}.

\begin{figure*}
\centering
{
  \includegraphics[width=0.48\textwidth]{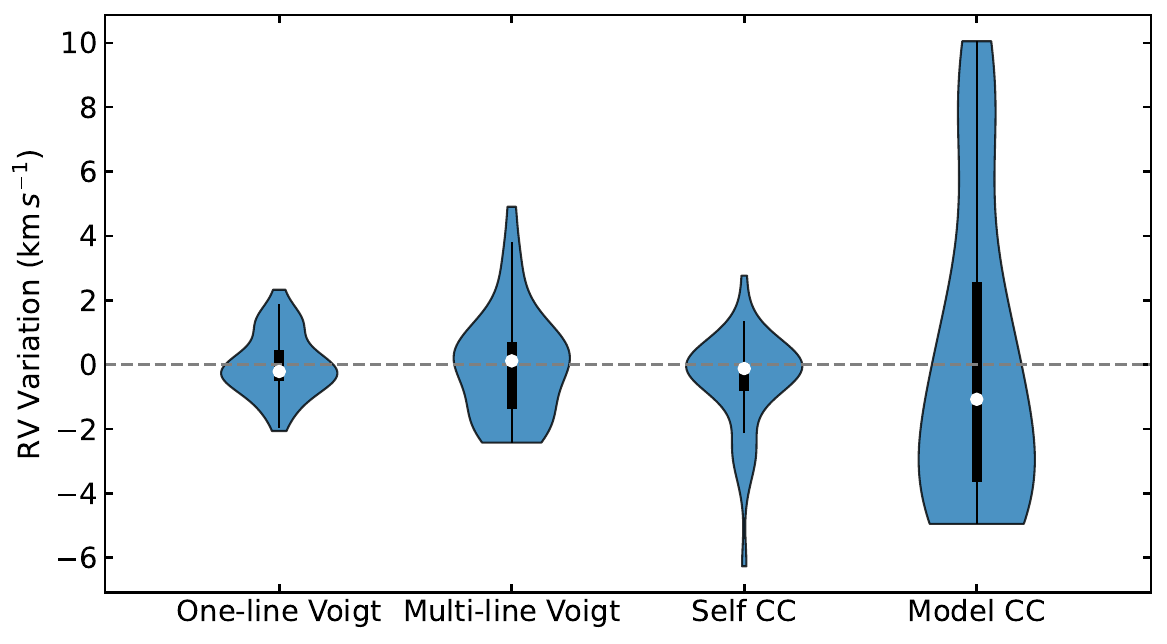}
}
\hspace{2mm}
{
  \includegraphics[width=0.48\textwidth]{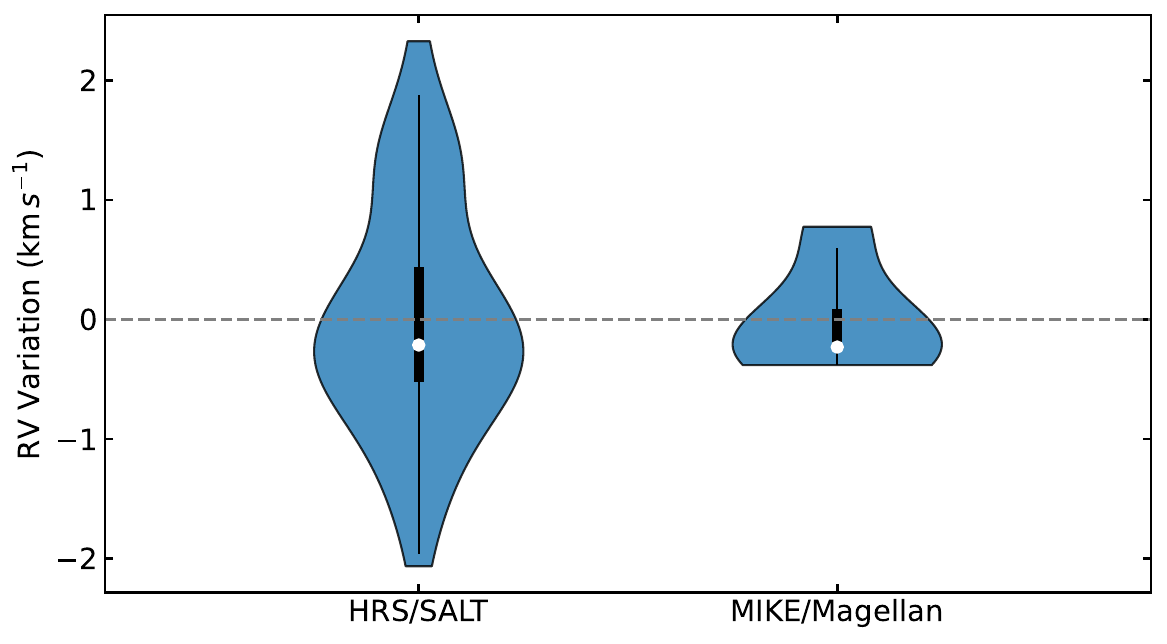}
}
\caption{Violin plots showing the distribution of radial velocity (RV) variations. (left) Comparison of the four methods tested on the HRS/SALT data: a single-line Voigt profile fitting (using the Mg\,\textsc{ii} 4481\AA\ line), multi-line Voigt profile fitting (using some of the strongest lines in the data: \ion{Mg}{2} 4481, \ion{Si}{2} 5055, 6347, 6371\,\AA\ lines), self cross-correlation (CC) where the highest SNR dataset was used as the template, and model cross-correlation where a white dwarf model was used for the template. For each method, the mean RV has been subtracted to highlight relative variations. White points indicate the median, thick black bars show the inter-quartile range, and the thin lines denote 1.5 times the inter-quartile range. (right) Comparison between the radial velocities derived from single-line Voigt profiles from HRS/SALT and MIKE/Magellan respectively showing MIKE/Magellan is more sensitive.}
\label{fig:SALT_MIKE_RV}
\end{figure*}

Different methods of radial velocity extraction were tested, including (i) single-line Voigt profile fitting (as outlined in Section~\ref{Analysis}), (ii) simultaneous multi-line Voigt profile fitting, (iii) cross-correlation techniques using the highest SNR spectrum as the master template, and (iv) cross-correlation techniques using synthetic template matching with white dwarf models. Among these, direct fitting of the individual strongest spectral line (most often Mg\,\textsc{ii} 4481\AA) yielded the lowest radial velocity scatter as shown in Fig.\,\ref{fig:SALT_MIKE_RV} (left). This likely reflects the wide wavelength separation of the spectral lines, which fall on different regions of the detector that are differentially affected by instrumental drifts. Therefore, methods which assume a coherent velocity shift across the full spectrum, such as cross-correlation or multi-line fitting are more susceptible to systematic errors in non-stabilized spectrographs. 

Adopting the direct Voigt profile fitting of the strongest metal line, the radial velocities were derived from the posterior distributions of the fitted line centroids of the Voigt profiles as reported in Section \ref{results-m-s} for each epoch, with uncertainties derived from the corresponding 16th and 84th percentiles on the line center. These line centers were converted into radial velocities and the null hypothesis was tested as to whether the radial velocities were consistent with a constant value. The $\tilde{\chi}^2$ statistic was used as described in Section \ref{results-m-s}. All targets yield a $\tilde{\chi}^2$ value below the 99.9\% critical threshold, indicating that the observed radial velocity variations are statistically consistent with no intrinsic variability, and so no real astrophysical signal is found. 

The measured 1\,$\sigma$ standard deviation of the radial velocities are 2.8\,km\,$s^{-1}$ (WD\,0106$-$328), 3.1\,km\,$s^{-1}$ (WD\,0408$-$041), 3.5\,km\,$s^{-1}$ (WD\,1457$-$086), 0.9\,km\,$s^{-1}$ (WD\,1929+011), and 3.1\,km\,$s^{-1}$ (G29-38). Although these values exceed the expected instrumental stability, the radial velocities are consistent with a flat line, and therefore these values are dominated by observational uncertainties rather than astrophysical signals. Interpreted as upper limits on the radial velocity semi-amplitudes, these constraints allow the exclusion of close-in stellar companions on inclined orbits (those that induce km\,$s^{-1}$ level signals) for all targets. For WD\,1929+011, the tighter constraint of 0.9\,km\,$s^{-1}$ additionally permits limits to be placed on massive close-in planetary companions, with companions with $M \sin (i)$ greater than a few Jupiter masses being detectable and therefore, disfavored in this system. 

Separating the datasets by instrument highlights the impact of data quality, with radial velocity scatter derived from MIKE/Magellan spectra being up to five times lower than that from SALT/HRS, as shown in Fig.\,\ref{fig:SALT_MIKE_RV} (right). This improvement reflects the combined effects of higher spectral resolution, increased SNR and improved instrument stability, highlighting that these factors dominate the limitations in the present analysis. 

\section{Discussion} \label{Discussion}

\subsection{Variability in metal lines}

This work aimed to investigate whether the mass of the pollutants in the photosphere of five warm DAZ white dwarfs changed over time. Given the dynamic nature of white dwarf planetary systems, variability in the accretion rate and therefore the amount of material present in the photosphere is expected. Over the baseline of up to 18 years, corresponding to hundreds to thousands of diffusion timescales for these stars, no statistically significant variability in four of the five stars was found. For one of the white dwarfs, WD\,0106$-$328, the ground-based data suggests variability in the equivalent width of the \ion{Mg}{2} doublet across 18 years, however, follow-up observations with the \textit{HST} show no statistically significant variability when comparing two ultraviolet spectra with more transitions separated by 9 years. These results therefore suggest that, for the majority of the systems studied here, the accretion of planetary material onto the white dwarf photosphere is stable on decadal timescales with inferred accretion rates being stable at the 15--30\% level (depending on the white dwarf). 

Polluted white dwarf systems are dynamic, with variability observed in the circumstellar material from as short as hours \citep[e.g.][]{Noor2025Activity} to as long as decades \citep[e.g.][]{farihi2018dust}. If the accretion is from asteroids with a range of sizes, then the observed pollution may be most typically dominated by the accretion of smaller asteroids that are scattered in often  \citep{wyatt2014stochastic,trierweiler2022exomoons}. However, the inferred accretion rates can increase following the scattering in of a larger asteroid, subsequently decreasing on a timescale determined by how long it takes for that material to be processed through the accretion disk to end up on the star, or by direct impacts onto the white dwarf's surface \citep{mcdonald2021white}. The lack of variability found in this study implies that scattering events that lead to variable accretion occur on timescales longer than decades or that accretion disks act to stabilize the accretion flow. The viscous timescale of a circumstellar accretion disk scales inversely with the viscosity parameter, $\alpha$, with the viscous timescale being tens of years if $\alpha$ is high ($\sim0.5$) and thousands of years if $\alpha$ is low ($\sim10^{-3}$) \citep{rafikov2011runaway}, suggesting that the stability of the accretion rates at 15--30\% over 15--18 years may favor a low-$\alpha$ regime, in which the disk viscous timescale is long compared to the observational baseline and acts to buffer short-term variations in the supply of material. In contrast, models invoking runaway gas-dust coupling instabilities or direct large-body impacts predict the possibility of order-of-magnitude transient excursions in the accretion rate \citep{mcdonald2021white, brouwers2022road, steckloff2021sublimation}. The present monitoring campaign cannot exclude short-duration outbursts occurring between observing epochs, but it does limit sustained long-term changes in the bulk accretion rate. The \textit{HST} observations of WD\,0106$-$328 constrain abundance changes between 2016 and 2025 to the $\lesssim$0.1 dex level, disfavoring a long-lived high-state accretion episode across that interval. It is worth noting that the probability of observing a rise in accretion rate from a recent accretion event increases with both the observational baseline and/or by increasing the sample of stars being monitored. The 15--30\% constraints on accretion rate stability over 18 years are a measure of the sensitivity to the disk processing timescale, and can be further improved by having a longer time baseline or higher equivalent width accuracy. 

\subsubsection{WD\,0106$-$328}

For WD\,0106$-$328, the ground-based spectra suggest variability in the equivalent width of the \ion{Mg}{2} doublet between the MIKE/Magellan and HRS/SALT observations. To assess the impact of differing spectral resolutions, the stacked MIKE/Magellan spectrum was interpolated and convolved to the same resolution as HRS/SALT (R$\sim$14,000). A significant difference in the line profiles persists between the MIKE/Magellan and early epochs of HRS/SALT (pre-2022) stacked spectra (Fig.\,\ref{fig:0106-stacked-res}) indicating that resolution alone cannot account for the discrepancy. When stacking all the data together for MIKE/Magellan and then for HRS/SALT, there is a 2.8\,$\sigma$ increase in the equivalent width between MIKE/Magellan and HRS/SALT, however, if only the early epochs of HRS/SALT are used (as shown in Fig.\,\ref{fig:0106-stacked-res}), this significance is instead 4.4\,$\sigma$. The stacked data show additional metal lines due to the higher SNR. The \ion{Fe}{2} 5018.44 and 5169.03\,\AA\ lines do not show a  change between these stacked datasets (Table \ref{tab:Multiple-Lines}), however, these lines are weak and given these data are stacked, the variability, if episodic, would likely be averaged over. WD\,0106$-$328 has particularly narrow metal lines in comparison to other DAZ white dwarfs, and the instrumental resolution of HRS/SALT is comparable to the intrinsic line width, meaning unresolved line structure may affect the measured equivalent widths. The absence of comparable discrepancies in the other four white dwarfs between MIKE/Magellan and HRS/SALT argues against systematic differences between the instruments.  

The lack of variability in the \textit{HST} spectra comparing data taken in 2016 and 2025 suggests that maintained long-term changes in bulk accretion rate are not occurring across the baseline of these observations. It remains possible that the ground-based data did not show true variability and agree with the \textit{HST} data given observational errors, resolution issues, or seeing constraints, however, it is also possible that there is a stable baseline accretion rate with episodic accretion above this level, rather than a sustained change in the bulk accretion rate. Figure \ref{fig:Comp-Farihi} compares the equivalent widths measured for the \ion{Mg}{2} 4481\AA\ and \ion{Ca}{2} 3933\AA\ K line derived in this work and those reported in \citet{Farihi2026accretion}. It should be noted that multiple spectrographs and a different method of deriving equivalent width were used, nonetheless, any equivalent width measurements made within a few weeks are consistent within 1$\sigma$. Furthermore, including these results in the statistics, the $\tilde{\chi}^2$ value increases to 5.29 for Mg and 5.36 for Ca compared against the critical  $\tilde{\chi}^2$ value of 2.08 and 2.84 respectively assuming a 99.9\% significance threshold, therefore, increasing the significance of the equivalent width variability. \citet{Farihi2026accretion} also reported temporal changes in the Ca/Mg ratio, however, no statistically significant ($>3\sigma$) change in Ca/Mg is found here, noting that the HRS/SALT data lack sensitivity in the blue.  

The equivalent widths can be translated into metal abundances and corresponding accretion rates (Table \ref{tab:WD-Acc-Rate}).  WD\,0106$-$328 exhibits a variable infrared excess attributed to a circumstellar dust disk \citep{swan2020dust,Noor2025Activity}. The mid-infrared flux is significantly decreasing across the $\sim$\,10 year baseline between the MIKE/Magellan and HRS/SALT observations. If the increase in equivalent width over time is real, this may provide evidence for the inner disk quickly evaporating and accreting such that there is anti-correlation between the infrared flux, and the accretion rate, as predicted in \citet{xu2014drop}. Using the maximum and minimum infrared fluxes reported by \citet{swan2020dust} and reported here in Table~\ref{tab:WD-IR-Var}, and assuming optically thin emission from micron-sized silicate grains at 1000\,K, the observed flux variation corresponds to a dust mass change of 6.8$\times 10^{16}$\,g (approximately 0.31 times the mass of comet Halley). This exceeds the spectroscopic detection threshold for changes in the accreted mass ($_{-0.04}^{+0.05}$ times the mass of comet Halley), implying that if this material were accreted efficiently on short timescales, a larger variation in the photospheric metal abundances would be detected. The absence of such variability indicates that the infrared variations do not directly trace changes in the instantaneous accretion rate. Instead the infrared variability likely reflects changes in the temperature structure, emitting area, or small-grain population that does not trace the bulk of the mass of the disk. The gaseous disk of material accreting onto the white dwarf likely regulates the accretion flow with potential metal line changes from episodic accretion. 

\subsubsection{WD\,0408$-$041}

No statistically significant variability is detected in the equivalent width of the \ion{Mg}{2} and Ca K lines for WD\,0408$-$041, which places an average 1\,$\sigma$ upper limit on accretion rate changes of 29\,\%. This corresponds to a 1\,$\sigma$ change in the accreted mass of  0.085 times the mass of comet Halley, implying that variations at or below this level would not produce a detectable change in the observed equivalent widths. The accretion rate onto this system is therefore stable within these limits over thousands of diffusion timescales. This is comparable to the results reported in \citet{farihi2018dust} where no variation in metal lines was reported.

Over a comparable baseline to these observations, infrared variability has been reported for WD\,0408$-$041 \citep{farihi2018dust,swan2020dust,Noor2025Activity}. If interpreted as a change in the mass of circumstellar dust, this would correspond to an additional mass of 7.05 times that of comet Halley, under the assumption of optically thin micron sized silicate grains. As with WD\,0106$-$328 this is above the detection threshold of variability from the spectroscopic monitoring again implying that the infrared variability does not directly trace the changes in the mass accretion rate onto this white dwarf, under the assumptions listed above.

\subsubsection{WD\,1457$-$086}

No statistically significant variations were detected in the metal lines for WD\,1457$-$086 which gave a 1\,$\sigma$ average accretion rate limit of 17\,\%, one of the tightest constraints of any system in this study. This is likely biased due to this white dwarf having the fewest epochs which pass the SNR threshold for the analysis. Therefore, over the duration of the observations, no change in mass accreted above 0.04 times the mass of comet Halley occurred over the total baseline of the observations. 

Although WD\,1457$-$086 is accreting at a high rate, there is no confirmed circumstellar disk \citep{dennihy2017wired}. WD\,1457$-$086 is the hottest white dwarf in the sample and it is plausible that this white dwarf could be accreting from a pure gas disk due to the sublimation of optically thin dust within the tidal disruption radius of the white dwarf \citep{bonsor2017infrared,steckloff2021sublimation}.

\subsubsection{WD\,1929+011}

WD\,1929+011 is a heavily polluted white dwarf with strong lines from numerous elements (Mg, Ca, Si, Fe, O) in the optical spectral range. No statistically significant variability was detected for WD\,1929+011 resulting in an average 1\,$\sigma$ accretion rate limit of 15\%. This has the strongest \ion{Mg}{2} line and Mg abundance of any white dwarf in the sample, meaning the resulting mass constraints are less stringent. No changes in mass accretion corresponding to mass above comet Halley sized objects was observed. 

WD\,1929+011 appears to have no variability either in the circumstellar dust nor the accretion rate. If changes in accretion rate are caused by new bodies being accreted, this implies that there are no recent, larger than comet Halley sized, collisional or tidal disruption events. Therefore, the processes driving the accretion of the circumstellar material are particularly stable in both the circumstellar environment and accretion mechanisms. 

\subsubsection{G29-38}

Searches for equivalent width changes for G29-38 have been studied previously by \citet{von2007discovery} and \citet{debes2008second}. The first data sets for G29-38 from MIKE/Magellan are the same as those reported in \citet{debes2008second}, allowing a direct comparison between methods to determine equivalent widths. The equivalent widths derived here for these data sets are consistent within 1\,$\sigma$ to those reported in \citet{debes2008second}. Small discrepancies between measurements may arise due to differing handling of the continuum normalization or the equivalent width measurement. \citet{debes2008second} normalized the regions around the spectral lines using high-order polynomials between third and twelfth order. Whereas, this work fitted lower order polynomials (3--5), and then allowed the continuum to be a free parameter in the Voigt model. The agreement between the two approaches provides confidence in the robustness of the equivalent width measurements. 

\citet{von2007discovery} reported variability with the \ion{Ca}{2} K line changing from 165\,m\AA\ to 280\,m\AA. Whereas, \citet{debes2008second} found the equivalent widths to be consistent with a median of 264\,m\AA\ and proposed that the previously reported variability likely arose from comparisons between low-resolution (R$\sim$500) and higher resolution spectra. In this work the best fitting equivalent width across the MIKE/Magellan observations was 267\,m\AA\ for the \ion{Ca}{2} K line, with no statistically significant variations found. This result supports the conclusion of \citet{debes2008second} that the metal absorption lines in G29-38 are stable over the timescales probed by these observations. It should be noted that these conclusions are based on a limited number of epochs, and episodic variability between observations remains possible. Furthermore, the reduced chi-squared statistic, $\tilde{\chi}^2$, was compared against the critical  $\tilde{\chi}^2$ value assuming a 99.9\% significance threshold. The measured $\tilde{\chi}^2$ value of 1.22 for the \ion{Mg}{2} line corresponds to a confidence level of approximately 83\%. Although this does not constitute a statistically significant detection of variability, it could still be possible. 

\subsection{Limitations and future work}

\subsubsection{Methods}

Equivalent width measurements provide a convenient tracer of line strength, however, they do not uniquely determine changes in elemental abundances. Variations in equivalent widths can arise from differences in continuum normalization, spectral resolution, or line blending, particularly in spectra with lower SNR. The spectra used in this analysis were obtained with multiple instruments, including MIKE/Magellan and SALT/HRS, which have different spectral resolutions, instrumental line spread functions, and wavelength ranges. Although care was taken when comparing data between instruments, subtle systematic differences may introduce  offsets in equivalent width measurements. Therefore, the strongest constraints on equivalent width variation come from comparisons of high SNR data taken on the same instrument. 

Fitting Voigt profiles using MCMC methods allows a full exploration of the posterior distribution of the Voigt parameters. However, for weaker lines and spectra with lower SNR, the MCMC did not converge or found unrealistic line profiles. This meant that low SNR data-sets (SNR$<$10) were omitted from the analysis and future observations should focus on higher SNR data. The results were also checked against direct line fitting, as outlined in Section~\ref{Analysis}, with no significant changes in the conclusions drawn. 

The temporal sampling of the spectroscopic data remains inherently sparse, with observations separated by intervals ranging from days to years. Such cadence limits sensitivity to variability occurring on short timescales, particularly those comparable to or shorter than the diffusion timescales of metals in DAZ white dwarfs. Transient accretion events or stochastic fluctuations in accretion rate may occur between observations and remain undetected resulting in an apparent stability in the measured equivalent widths. In addition, sparse sampling reduces the ability to characterize the amplitude and period of any variability and may bias measurements toward long-term averages rather than instantaneous states. Therefore, the absence of observed variability here should be interpreted as a constraint on sustained or large-amplitude changes, rather than ruling out short-lived or rapidly evolving accretion episodes. High-cadence, time-resolved spectroscopic monitoring is therefore required to fully capture the dynamical behavior of metal accretion in these systems. 

\begin{table*}
	\centering
	\footnotesize
	\caption{The $\tilde{\chi}^2$ and $\chi ^2_{\textrm{crit}}$ values for the Mg equivalent width measurements, together with the best-fitting equivalent width (EW) and its standard deviation. These are converted into Mg abundance and associated abundance error. The mass accretion rate is calculated assuming a bulk Earth composition using the best-fitting [Mg/H] abundance $\dot{M} = (100/15.4) \times M_{\textrm{WD}} \times 10^q \times 10^{[\textrm{Mg}/\textrm{H(e)}]} \times A_{\textrm{Mg/H}} / \tau_{\textrm{Mg}}$, where $q = \log _{10}(M_{\textrm{CVZ}}/ M_{\textrm{WD}})$, $A_{\textrm{Mg/H}}$ is the atomic mass of Mg to  H, and $\tau_{\textrm{Mg}}$ is the Mg sinking timescale (Table \ref{tab:WD-sample}). ${M}_{\textrm{total}}$ denotes the total mass accreted over the full observational baseline. Assuming the photospheric abundance varies by $\pm\,1\sigma$ from the best-fitting value, $\Delta M_{\textrm{acc,T}}$ gives the corresponding change in accreted mass over this baseline, expressed as a fraction of the mass of Comet Halley (2.2$\times 10 ^{17}$\,g).}
	\label{tab:WD-Acc-Rate}
	\begin{tabular}{cccccccccc} 
		\hline
        WD & $\tilde{\chi}^2$ & $\chi ^2_{\textrm{crit}}$ & EW & [Mg/H] & $\dot{M}$  & ${M}_{\textrm{total}}$  & $\Delta M_{\textrm{acc,T}}$ \\
         &  &  &  (m\AA) &  &  (g\,s$^{-1}$) &  (g) & ($M_{\textrm{C. Halley}}$)  \\
		\hline	

        \vspace{1mm} WD0106$-$328 & 4.73 & 2.27 & 110 \err 26 & $-$5.67 $_{-0.15}^{+0.13}$ & 5.7 $_{-1.7}^{+2.1}\times 10^{7}$ & 3.2 $_{-0.9}^{+1.2}\times 10^{16}$ &  $_{-0.04}^{+0.05}$ \\ 
 
        \vspace{1mm} WD0408$-$041 & 1.79 & 2.35 & 162 \err 30 & $-$5.45 $_{-0.14}^{+0.11}$ & 1.2 $_{-0.3}^{+0.4}\times 10^{8}$ & 6.6 $_{-1.8}^{+1.9}\times 10^{16}$ &  $_{-0.08}^{+0.09}$ \\ 
       
        \vspace{1mm} WD1457$-$086 & 0.68 & 3.10 & 91 \err 11 & $-$5.53 $_{-0.08}^{+0.07}$ & 1.2 $_{-0.2}^{+0.2}\times 10^{8}$ & 5.9 $_{-1.0}^{+1.0}\times 10^{16}$ &  $_{-0.04}^{+0.04}$ \\ 
        
        \vspace{1mm} WD1929+011 & 1.10 & 1.82 & 468 \err 26 & $-$4.14 $_{-0.07}^{+0.06}$ & 2.0 $_{-0.3}^{+0.3}\times 10^{9}$ & 9.6 $_{-1.4}^{+1.4}\times 10^{17}$ &  $_{-0.62}^{+0.65}$ \\ 

        \vspace{1mm} G29-38 & 1.22 & 1.87 & 43 \err 7 & $-$5.91 $_{-0.08}^{+0.07}$ & 4.4 $_{-0.7}^{+0.8}\times 10^{8}$ & 2.5 $_{-0.4}^{+0.4} \times 10^{17}$ &  $_{-0.19}^{+0.20}$ \\ 

		\hline

	\end{tabular}
\end{table*}

\subsubsection{White dwarf models}

Models of elemental sedimentation based on one-dimensional (1D) mixing-length theory predicts that the settling timescales are on the order of days for warm ($\gtrsim$15\,000\,K) DAZ white dwarfs \citep{koester2009accretion,koester2020new}.
Understanding mixing mechanisms including convection and thermohaline mixing in the atmospheres of white dwarfs is crucial to ensure accurate sinking timescales. \citet{cunningham2019convective} performed three-dimensional (3D) radiation-hydrodynamic simulations that investigated the role of convective overshoot in determining accurate timescales of elemental sedimentation for white dwarfs with convective atmospheres. The onset of substantial convective instabilities occurs at 18,000--18,250\,K, DA white dwarfs cooler than this may therefore be unstable to convection which consequently may increase the diffusion timescales by 1.5--3 orders of magnitude  \citep{cunningham2019convective}. WD\,0106$-$328, WD\,0408$-$041, and G29-38 lie in this temperature regime and may therefore have longer sinking timescales than those reported in Table~\ref{tab:WD-sample}. For WD\,0106$-$328 the sinking timescale increases to 4.0 days for Ca and 6.4 days for Mg (factor 3.5 increase), for WD\,0408$-$041 it increases to 33.3 and 37.6 days for Ca and Mg respectively (factor 8 increase), and G29-38 increases to 658.3 and 549.2 days for Ca and Mg respectively (factor 8 increase) \citep{cunningham2019convective}. Therefore, especially for G29-38, these observations cover fewer sinking timescales, and further follow up spectra would enable more precise constraints on variability. 

It has been hypothesized that DAZ white dwarfs ($>$\,10,500\,K) do not spread the metals homogeneously across the surface, therefore predicting surface abundance variations of the metals, if the metals are accreted with a heterogeneous surface distribution \citep{cunningham2021horizontal}. The rotation periods of white dwarfs are longer than the spectroscopic exposure times in this work, and therefore, this work should have been sensitive to surface feature changes, providing the abundance variations would be significant enough to detect. Therefore, this may imply that the accretion is spherically homogeneous. 

Models of thermohaline mixing instabilities have been used to predict that the mixed mass in white dwarf atmospheres may be orders of magnitude larger and have been used to offer an explanation for  the orders of magnitude lower accretion rates of DAs than DBs \citep{bauer2018increases,bauer2019polluted,Buchan2025Exogeological}. This enhanced mixing can also affect the vertical sinking timescales. Thermohaline mixing is driven by an inverted molecular weight gradient, which could be established via the accretion of metals. If accretion stops, the sudden absence of the inverted chemical gradient is expected to render thermohaline mixing irrelevant \citep{bauer2019polluted}, leading to diffusion timescales governed by gravitational sedimentation, or a combination of the latter and convection if the atmosphere is unstable to convection. 

Existing models of white dwarf atmospheric mixing do not predict sinking timescales long enough, for the DAZs in this work, to smooth out significant variations in the accretion rate over the observational baselines used in this study. The stability of the inferred accretion rates to within 15--30\% over 15--18 years therefore most plausibly reflects genuinely stable accretion from a long-lived circumstellar reservoir, rather than a photospheric averaging effect. Continued refinement of mixing models will enable further understanding as to whether variability in the metal abundances in the photospheres are expected. Model independent measurements of instantaneous accretion rates via X-ray observations \citep{Cunningham2022Xrays} offer a complementary route to disentangling accretion variability from photospheric processing.

\subsubsection{Future observations}

Continued spectroscopic monitoring of polluted white dwarfs provides a powerful means of probing stability and variability of accretion from remnant planetary systems. The sensitivity achieved in this work demonstrated that such observations are capable of detecting changes in accreted mass on the scale of individual minor bodies, comparable down to a few per cent of the mass of comet Halley. This highlights the potential of long-term monitoring campaigns to directly trace the delivery and evolution of planetary bodies around white dwarfs. 

These results underscore the importance of observational homogeneity as subtle systematic differences between instruments, particularly in spectral resolution, can mimic or obscure intrinsic variability at the level of interest. Future programs should therefore prioritize repeated observations with the same instrumental configuration with higher SNR ($>10$), or ensure rigorous cross-calibration between datasets in order to robustly detect and interpret small-amplitude changes in metal line strengths. 

For radial velocity derivations, more accurate and precise measurements require stable and temperature controlled spectrographs. The metal lines in polluted white dwarfs are sufficiently narrow that planetary mass objects are possible to detect using the radial velocity method and therefore, future work should prioritize time series observations using higher precision stable spectrographs. 

\begin{table*}
	\centering
	\footnotesize
	\caption{Infrared variability from previous studies. The \textit{K} band data is from \citet{rogers2020near} where the variability measure is an upper limit that corresponds to the measurement limit of the observations. The Spitzer data are from \citet{swan2020dust} showing the percentage change, significance of the change, and the lowest and highest flux values for the 4.5\,$\micron$ channel. M$_{\textrm{dust}}$ gives the mass of optically thin dust that gives the infrared excess at the lowest and highest Spitzer 4.5\,$\micron$ fluxes assuming 1\,$\micron$ sized astronomical silicate grains \citep{Laor1993SilicatesKappa}. $\Delta$M$_{\textrm{dust}}$ is the difference in micron sized dust that would cause this difference in infrared flux. $\tau _{\textrm{dust}}$ is how long the disk would survive given the derived accretion rate in Table \ref{tab:WD-Acc-Rate}. WD\,1457$-$086 is excluded as it does not have a confirmed infrared excess from a dust disk. }
	\label{tab:WD-IR-Var}
	\begin{tabular}{cccccccccccc} 
		\hline
		WD Name & K & 3.6\,$\micron$ & 4.5\,$\micron$ & F$_{\mathrm{4.5,low}}$ & F$_{\mathrm{4.5,high}}$ & M$_{\textrm{dust,low}}$ & M$_{\textrm{dust,high}}$ & $\Delta$M$_{\textrm{dust}}$ & $\Delta$M$_{\textrm{dust}}$ & $\tau _{\textrm{dust}}$\\
         & (\%) & (\%) & (\%) & ($\mu$Jy) & ($\mu$Jy) & (g) & (g) & (g) & ($M_{\textrm{C. Halley}}$) & (yrs) \\
		\hline	
WD\,0106$-$328 & $<$3.9 & 9.0 (6$\sigma$) & 10 (6$\sigma$) & 137 & 150 & 1.63E+17 & 2.31E+17 & 6.8E+16 & 0.31 & 90--129 \\
WD\,0408$-$041 & $<$1.1 & 16 (9$\sigma$) & 26 (13$\sigma$) & 1099 & 1381 & 5.54E+18 & 7.09E+18 & 1.55E+18 & 7.05 & 1464--1874 \\
WD\,1929+011 & $<$0.64 & 1.0 (0.7$\sigma$) & 2.0 (2$\sigma$) & 897 & 920 & 1.83E+18 & 1.90E+18 & 7.00E+16 & 0.32 & 29--30\\
G29-38 & $<$1.0 & 12 (7$\sigma$) & 9.0 (4$\sigma$) & 8801 & 9559 & 2.66E+18 & 2.90E+18 & 2.4E+17 & 1.09 & 192--209 \\
		\hline
	\end{tabular}
    \textbf{ \ \\ References:} (1) \citet{xu2019compositions}, (2) \citet{melis2011accretion}, (3) \citet{xu2014elemental}.
\end{table*}

\section{Conclusions} \label{Conclusions}

This work presents a long-term spectroscopic monitoring campaign of five DAZ white dwarfs, spanning baselines in excess of 15 years with 18--47 epochs per object and sampling timescales from days to decades. These observations probe hundreds to thousands of diffusion (sinking) timescales based on standard 1D atmospheric models and therefore provide stringent constraints on sustained variability in photospheric metal abundances and accretion rates,  with accretion rates found to be stable to within 15--30\% and can probe down to a few per cent the mass of comet Halley. 

Equivalent widths of the dominant metal absorption lines were measured for each system. For four of the five white dwarfs in the sample, no statistically significant variability was detected across the full observational baseline. For WD\,0106$-$328, ground-based spectra taken between 2007 and 2025 indicate statistically significant variability in the \ion{Mg}{2} doublet over the 18 year monitoring program, however, a comparison between two \textit{HST} spectra taken in 2016 and 2025 found no statistically significant difference in line strengths. If the variability inferred from the ground-based data is intrinsic, the system must have undergone temporal changes in accretion that subsequently returned to a comparable photospheric abundance level. These results highlight the importance of homogeneous observing strategies, as systematic differences between instruments, particularly in spectral resolution, can introduce uncertainties comparable to the variability being probed. 

The absence of detectable variability in the majority of the sample indicates that the underlying source of the accretion is broadly stable at the level of sensitivity achieved over decadal timescales. This places constraints on the physics of circumstellar accretion, and supports a scenario where a viscous gas accretion disk smooths stochastic delivery of material from disrupted planetesimals.

These findings highlight that time-domain spectroscopy of polluted white dwarfs can provide a powerful probe of accretion physics. The sensitivity achieved is sufficient to constrain changes in accreted mass at the level of minor bodies, highlighting the potential of such observations to directly trace the delivery of planetary debris. Future monitoring campaigns, particularly those with homogeneous, high-cadence observations, will be essential for resolving short-timescale variability, refining constraints on diffusion physics, and establishing the physical connection between circumstellar disk evolution and accretion onto the white dwarf surface. 

\begin{acknowledgments}

LKR is supported by NOIRLab, which is managed by the Association of Universities for Research in Astronomy (AURA) under a cooperative agreement with the U.S. National Science Foundation. SX is supported by the international Gemini Observatory, a program of NSF NOIRLab, which is managed by the Association of Universities for Research in Astronomy (AURA) under a cooperative agreement with the U.S. National Science Foundation, on behalf of the Gemini partnership of Argentina, Brazil, Canada, Chile, the Republic of Korea, and the United States of America. AS acknowledges receiving funding from the European Research Council (ERC) under the European Union's Horizon 2020 research and innovation programme (Grant agreement No. 101020057). MCW was supported by the Science and Technology Facilities Council grant UKRI1198. TC was supported by NASA through the NASA Hubble Fellowship grant HST-HF2-51527.001-A awarded by the Space Telescope Science Institute, which is operated by the Association of Universities for Research in Astronomy, Inc., for NASA, under contract NAS5-26555.

This paper includes data gathered with the 6.5 meter Magellan Telescopes located at Las Campanas Observatory, Chile. Some of the observations reported in this paper were obtained with the Southern African Large Telescope (SALT). This research is based on observations made with the NASA/ESA Hubble Space Telescope obtained from the Space Telescope Science Institute, which is operated by the Association of Universities for Research in Astronomy, Inc., under NASA contract NAS 5–26555. These observations are associated with programs 14597 and 17819. This work has made use of data from the European Space Agency (ESA) mission {\it Gaia} (\url{https://www.cosmos.esa.int/gaia}), processed by the {\it Gaia} Data Processing and Analysis Consortium (DPAC, \url{https://www.cosmos.esa.int/web/gaia/dpac/consortium}). Funding for the DPAC has been provided by national institutions, in particular the institutions participating in the {\it Gaia} Multilateral Agreement. 

\end{acknowledgments}

\begin{contribution}

LKR led the data analysis, and was responsible for preparing and submitting the manuscript. MS contributed to developing the scientific concepts, wrote observing proposals, and oversaw data acquisition. AB and SX provided supervision and scientific guidance throughout the project. ELB and PD fitted white dwarf atmospheric models to optical and ultraviolet spectroscopic data. JD contributed to proposal writing,  data acquisition, and developed methodology for spectral fitting. ON fitted the spectra to derive RV constraints using multiple methods. TvH initiated the project and contributed to its scientific development. ED and MBA obtained, reduced, and calibrated MIKE/Magellan spectroscopic data. SH and MW provided expertise and guidance on data analysis and interpretation. AS measured infrared fluxes for the white dwarf sample and contributed to the analysis. TC developed and provided 3D atmospheric models for the white dwarfs. All authors reviewed and edited the manuscript.


\end{contribution}

%
\facilities{HST(COS), SALT(HRS), Magellan(MIKE)}

\software{COS notebooks, numpy, matplotlib, \textsc{scipy}\footnote{\url{https://scipy.org}}, \textsc{emcee}\footnote{\url{https://emcee.readthedocs.io/en/stable}}}


\appendix
\section{Appendix A - absorption lines}

Table \ref{tab:EW-HST-WD0106} reports the lines detected in the two sets of ultraviolet \textit{HST} data for WD\,0106$-$328. The table lists the line, transition information, and equivalent widths measured for the 2016 and 2025 data. 

\begin{deluxetable}{cccccc}
\label{tab:EW-HST-WD0106}
\tablecaption{Details on the lines detected in the \textit{HST} ultraviolet spectra for WD\,0106$-$328. Only lines detected at 3\,$\sigma$ in both the 2016 and 2025 spectra are listed.}
\tablehead{\colhead{Element} & $\lambda _{\textrm{\,vac}}$  (\AA) & log(gf) & E$_{low}$ (eV) & \colhead{EW 2016} (m\AA) & \colhead{EW 2025} (m\AA) }
\startdata
Si & 1142.29 & 0.011 & 16.098 & 43$_{- 11 }^{+ 12 }$ &  50$_{- 13 }^{+ 23 }$ \\
Fe & 1144.94 & 0.037 & 0.000 & 42$_{- 10 }^{+ 14 }$ &  48$_{- 9 }^{+ 10}$ \\
O & 1152.15 & $-$0.268 & 1.967 & 45$_{- 12 }^{+ 16 }$ &  46$_{- 7 }^{+ 8 }$ \\
Si & 1260.42 & 0.462 & 0.000 & 328$_{- 28 }^{+ 31 }$ &  318$_{- 29 }^{+ 51 }$ \\
Si & 1264.74 & 0.71 & 0.036 & 381$_{- 28 }^{+ 29 }$ &  326$_{- 36 }^{+ 45 }$ \\
Si & 1265.01  & $-$0.345 & 0.0356 & 100$ _{- 16 }^{+ 18 }$ &  159$ _{- 30 }^{+ 32 }$ \\
Si & 1294.55 & $-$0.037 & 6.553 & 62$ _{- 13 }^{+ 17 }$ &  31$ _{- 5 }^{+ 6 }$ \\
Si & 1296.73 & $-$0.127 & 6.537 & 37$ _{- 6 }^{+ 7 }$ &  39$ _{- 8 }^{+ 9 }$ \\
Si & 1298.89 & $-$0.257 & 6.553 & 106$ _{- 18 }^{+ 17 }$ &  74$ _{- 11 }^{+ 12 }$ \\
Si & 1301.15 & $-$0.127 & 6.553 & 57$ _{- 16 }^{+ 25 }$ &  48$ _{- 11 }^{+ 15 }$ \\
O & 1302.17 & $-$0.585 & 0.000 & 69$ _{- 8 }^{+ 9 }$ &  69$ _{- 9 }^{+ 13 }$ \\
Si & 1303.32 & $-$0.037 & 6.585 & 45$ _{- 8 }^{+ 11 }$ &  46$ _{- 10 }^{+ 18 }$ \\
Si & 1304.37 & $-$0.423 & 0.000 & 42$ _{- 7 }^{+ 7 }$ &  53$ _{- 6 }^{+ 7 }$ \\
O & 1304.86 & $-$0.808 & 0.020 & 31$ _{- 4 }^{+ 4 }$ &  29$ _{- 5 }^{+ 5 }$ \\
Si & 1305.59 & 6.859 & 0.710 & 46$ _{- 7 }^{+ 8 }$ &  80$ _{- 13 }^{+ 13 }$ \\
O & 1306.03 & $-$1.285 & 0.028 & 37$ _{- 6 }^{+ 7 }$ &  40$ _{- 7 }^{+ 11 }$ \\
Si & 1309.28 & $-$0.448 & 0.036 & 200$ _{- 20 }^{+ 20 }$ &  197$ _{- 15 }^{+ 18 }$ \\
Si & 1346.88 & $-$0.144 & 5.323 & 51$ _{- 8 }^{+ 10 }$ &  42$ _{- 7 }^{+ 9 }$ \\
Si & 1348.54 & $-$0.186 & 5.309 & 29$ _{- 6 }^{+ 7 }$ &  31$ _{- 5 }^{+ 6 }$ \\
Si & 1350.07 & 0.216 & 5.345 & 43$ _{- 5 }^{+ 6 }$ &  60$ _{- 7 }^{+ 8 }$ \\
Si & 1352.64 & $-$0.193 & 5.323 & 40$ _{- 7 }^{+ 9 }$ &  34$ _{- 9 }^{+ 10 }$ \\
Si & 1353.72 & $-$0.158 & 5.345 & 28$ _{- 6 }^{+ 9 }$ &  39$ _{- 8 }^{+ 9 }$ \\
\enddata
\end{deluxetable}

\section{Appendix B - WD\,0106$-$328 equivalent width variation}

\begin{figure*}
\centering
{
  \includegraphics[width=0.48\textwidth]{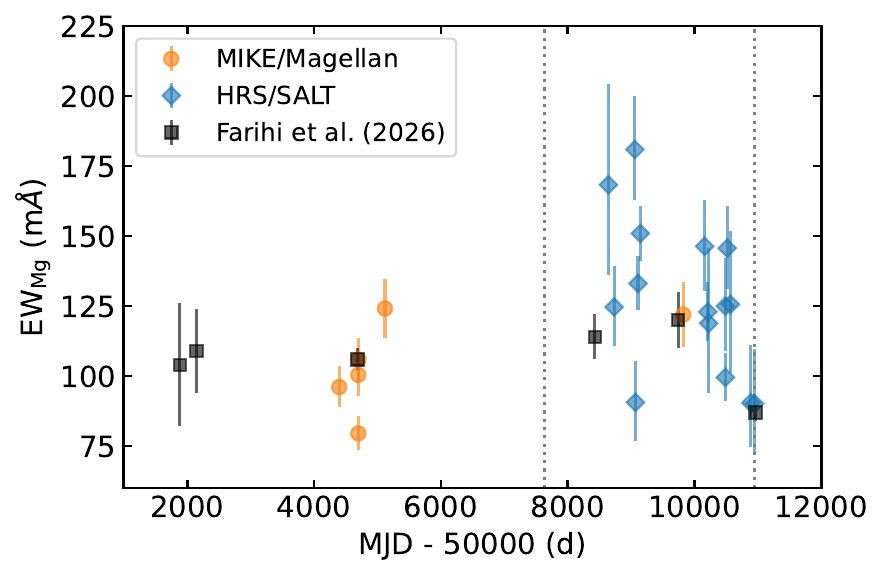}
}
\hspace{2mm}
{
  \includegraphics[width=0.48\textwidth]{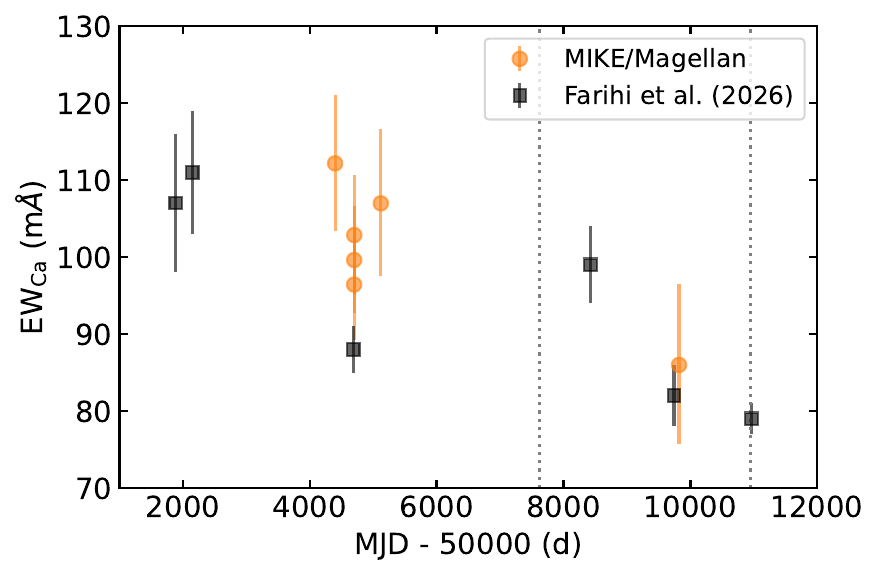}
}
\caption{A comparison between equivalent widths derived from this work (MIKE/Magellan and HRS/SALT) and \citet{Farihi2026accretion} using the (left) \ion{Mg}{2} 4481\AA\ line and (right) \ion{Ca}{2} 3933\AA\ K line for WD\,0106$-$328. All measurements taken within a few weeks of one another have equivalent widths consistent within 1\,$\sigma$.}
\label{fig:Comp-Farihi}
\end{figure*}

Figure \ref{fig:Comp-Farihi} compares the equivalent widths derived for (left) the \ion{Mg}{2} 4481\AA\ line and (right) the \ion{Ca}{2} 3933\AA\ K for this work (MIKE/Magellan and HRS/SALT) compared with those derived in \citet{Farihi2026accretion}. The closest in time observations were taken 13 days apart around 4700\,MJD-50000 days, and these equivalent widths, although derived with different methods, are consistent within 1\,$\sigma$ for both the Mg and Ca lines. 

\bibliography{Master-Bib}{}
\bibliographystyle{aasjournalv7}



\end{CJK}
\end{document}